\documentclass[a4paper,12pt]{article}
\usepackage{epsf,axodraw}
\usepackage{graphicx}
\usepackage{amsmath}

\newcommand{\lqcd}{\Lambda_{\textrm{\scriptsize{QCD}}}}
\newcommand{\la}{\langle}

\def\bq{\begin{quote}}
\def\eq{\end{quote}}

\def\bwg{{B\to\gamma\ell\nu_\ell}}
\def\dirac#1{#1\llap{/}}
\def\pv#1{\vec{#1}_\perp}
\def\as{\alpha_s}
\def\tp{\tilde{k}_+}
\def\eg{E_\gamma}

\begin{document}

\thispagestyle{empty}

\begin{flushright}
CERN-TH/2002-228\\ SHEP/02-16
\end{flushright}

\vspace{\baselineskip}

\begin{center}
\vspace{0.5\baselineskip} \textbf{\Large Factorization, the
Light-Cone Distribution \\[0.2em] Amplitude of the
\boldmath{$B$}-Meson\\[0.5em] and the Radiative Decay
\boldmath{$\bwg$}}\\ \vspace{3\baselineskip} {{\sc
S.~Descotes-Genon$^{\,a}$} and {\sc C.T.~Sachrajda$^{a,b}$}}\\
\vspace{2\baselineskip} \textit{$^{a}$ Department of Physics and
Astronomy, University of Southampton\\[0.1cm] Southampton, SO17
1BJ, U.K.\\[0.2cm] $^{b}$ Theory Division, CERN, CH-1211 Geneva
23, Switzerland} \\ \vspace{3\baselineskip}

\vspace*{0.8cm}
\textbf{Abstract}\\
\vspace{1\baselineskip}
\parbox{0.9\textwidth}{We study the radiative decay $\bwg$ in the
framework of \textit{QCD factorization}. We demonstrate explicitly
that, in the heavy-quark limit and at one-loop order in
perturbation theory, the amplitude does factorize, i.e. that it
can be written as a convolution of a perturbatively calculable
hard-scattering amplitude with the (non-perturbative) light-cone
distribution amplitude of the $B$-meson. We evaluate the
hard-scattering amplitude at one-loop order and verify that the
\textit{large logarithms} are those expected from a study of the
$b\to u$ transition in the \textit{Soft-Collinear Effective
Theory.} Assuming that this is also the case at higher orders, we
resum the large logarithms and perform an exploratory
phenomenological analysis. The questions addressed in this study
are also relevant for the applications of the QCD factorization
formalism to two-body non-leptonic $B$-decays, in particular to
the component of the amplitude arising from hard spectator
interactions.}
\end{center}

\newpage

\setcounter{page}{1}

%\tableofcontents

\newpage

\section{Introduction}\label{sec:intro}

The study of $B$-decays is providing a wonderful opportunity to
test and improve our understanding of the Standard Model of
Particle Physics and of its limitations. In particular, the BaBar
and Belle $B$-factories~\cite{babar,belle}, as well as other
experiments, are providing us with an impressive amount of
accurate experimental information about two-body non-leptonic
$B$-decays. Among the remarkable achievements is the increasingly
precise determination of $\sin(2\beta)$ (where $\beta$ is one of
the angles of the unitarity triangle) from measurements of the
mixing-induced CP-asymmetry in the golden mode, $B\to J/\psi
K_s$~\cite{sin2b}. Unfortunately the precision with which we can
determine the fundamental properties and parameters of the
standard model (in particular the CKM matrix elements) from the
measured branching ratios and asymmetries of other non-leptonic
channels is severely limited by our inability to control
non-perturbative QCD effects. Further progress in overcoming this
limitation is urgently needed if we are to be able to exploit
effectively the wealth of experimental data for fundamental
physics.

\begin{figure}[t]
   \vspace{-3cm}
   \epsfysize=23cm
   \epsfxsize=16cm
   \centerline{\epsffile{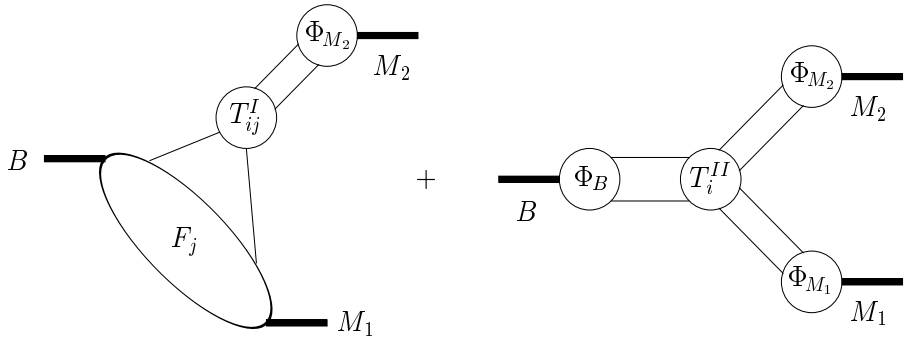}}
   \vspace*{-15.5cm}
\caption[dummy]{\label{fig1}\small Graphical representation of the
factorization formula. Only one of the two form-factor terms in
eq.~(\ref{fff}) is shown for simplicity.}
\end{figure}

An important step towards the control of non-perturbative QCD
effects in two-body non-leptonic $B$-decays has been the recent
discovery that in the heavy-quark limit, $m_b\to \infty$ (where
$m_b$ is the mass of the $b$-quark), hard and soft physics can be
separated (factorized)~\cite{BBNS1,BBNS2}. Within this
\textit{factorization} framework the amplitudes are expressed as
convolutions of perturbatively calculable hard-scattering kernels
and universal non-perturbative quantities (light-cone distribution
amplitudes and semi-leptonic form factors). When the decay
products are light mesons $M_{1,2}$, such as in $B\to\pi\pi$ or
$B\to\pi K$ decays, the factorization formulae take the generic
form, represented schematically in fig.\,\ref{fig1},
\begin{eqnarray}
\langle M_1 M_2|{\cal O}_i|\bar{B}\rangle &=& \sum_j F_j^{B\to
M_1}(m_2^2)\,\int_0^1 du\,T_{ij}^I(u)\,\Phi_{M_2}(u)
\,\,+\,\,(M_1\leftrightarrow M_2)\nonumber\\ &&\hspace*{-2cm}
+\,\int_0^1 d\xi du dv \,T_i^{II}(\xi,u,v)\,
\Phi_B(\xi)\,\Phi_{M_1}(v)\,\Phi_{M_2}(u)\,. \label{fff}
\end{eqnarray}
Here ${\cal O}_i$ is one of the operators of the effective weak
Hamiltonian, $F_j^{B\to M_{1,2}}$ denotes the $B\to M_{1,2}$ form
factors, and $\Phi_X(u)$ is the light-cone distribution amplitude
for the quark-antiquark Fock state of meson $X$. $T_{ij}^I(u)$ and
$T_i^{II}(\xi,u,v)$ are hard-scattering functions, which are
perturbatively calculable. The hard-scattering kernels and
light-cone distribution amplitudes depend on a factorization scale
and scheme; this dependence is suppressed in the notation of
eq.~(\ref{fff}). Finally, $m_{1,2}$ denote the light meson masses.
In phenomenological applications of the factorization framework
(see, for example refs.\,\cite{BBNS2,MN2002}), the hard-scattering
kernels have been computed to $O(\alpha_s)$. For the $T^I_{ij}$'s,
for which the leading contribution is generally of order
$\alpha_s^0$, this requires the evaluation of one-loop diagrams,
whereas for the $T^{II}_i$'s, which start at $O(\alpha_s)$, one
only needs to evaluate tree-level diagrams. An important
difference in the two cases is that the hard scale in the
$T^I_{ij}$'s is $m_b$, whereas in the $T^{II}_i$'s it is
$\sqrt{m_b\lqcd}$. The hard spectator interaction term in
eq.\,(\ref{fff}), i.e. the component containing the $T^{II}_i$'s,
therefore depends on three scales, $m_b$, $\sqrt{m_b\lqcd}$ and
$\lqcd$, and it is important to understand the structure of the
higher order corrections. Note that since we work at leading twist
we do not distinguish between the quark mass $m_b$, the meson mass
$M_B$ or any scale which differs from these by $O(\lqcd)$.

%%%%%%%%%%%%%%%%%%%%%%%%%%%%%%%%%%%%%%%%%%%%%%%%%%%%%%%%%%%%%%%%%%%
%%%%%%%%%%%%%%%%%%%%%%%%%%%%%%%%%%%%%%%%%%%%%%%%%%%%%%%%%%%%%%%%%%%

In this paper we study factorization, and in particular the
one-loop corrections, in a simpler process, the radiative decay
$\bwg$, where the photon is hard. This process is interesting in
itself, and also has the key features of the $T^{II}_i$ term in
eq.\,(\ref{fff}), in particular the dependence on the three
scales. The only hadron in the decay is the $B$-meson, making it
easier to focus on the properties of $\Phi_B$ and on issues
related to the presence of the three scales. The questions which
we investigate include:
\begin{enumerate}
\item Does factorization hold at one-loop order?

This has not yet been directly verified for the $T^{II}_i$ term in
eq.\,(\ref{fff}). Indeed it has been suggested that it is
necessary to replace the light-cone distribution amplitude of the
$B$-meson by a wave function which also depends on the transverse
components of momentum~\cite{KPY}.\label{item:fact}

With the definition of the $B$-meson's light-cone distribution
amplitude given in eq.~(\ref{eq:phihdef})
below~\cite{BBNS1,BF,GN}, we verify that factorization does hold
at one-loop order and that there is no need to introduce a
dependence on transverse components of momentum. We evaluate the
hard-scattering kernel explicitly.
\item How can one obtain information about the light-cone distribution
amplitude, $\Phi_B$, non-perturbatively, for example from lattice
simulations?

For light mesons such as the pion, moments of the distribution
amplitude are given in terms of matrix elements of local
operators, and can be determined using non-perturbative methods
such as lattice simulations or QCD sum rules. For the $B$-meson
the situation is different. The presence of ultraviolet
divergences in the integral over a light-cone component of
momentum implies that the corresponding matrix elements of local
operators do not give information useful for the decay amplitude.
The distribution amplitude will therefore have to be determined by
evaluating the matrix elements of non-local operators, which is
possible in principle, but considerably more difficult.
Alternatively, one may attempt to determine the distribution
amplitude from experimental measurements of processes such as the
one being studied in this paper. The additional ultraviolet
divergences also complicate the dependence of the distribution
amplitude on the factorization scale (i.e. the evolution). The
presence of such additional ultraviolet divergences has been
stressed previously in ref.~\cite{GN}, and questions concerning
the evolution of the distribution amplitude were discussed in
ref.~\cite{KPY}. \label{item:how}
\item How can one resum the large logarithms which appear in
one-loop perturbation theory?

In particular one has the Sudakov double logarithms associated
with the heavy-to-light ($b\to u$) decay. Large logarithms appear
both in the distribution amplitude and in the hard-scattering
kernel. In this context we find the formulation of the
soft-collinear effective theory (SCET)~\cite{SCET2,SCET1} very
helpful, and we present the resummed expressions in
sec.\,\ref{sec:resum}. We stress however, that, whereas we have
performed the one-loop calculation explicitly, the validity of the
all-orders resummation in the present context still requires a
formal demonstration.

For some other processes, for example for the semi-leptonic
$B\to\pi$ form factor at large momentum transfer, the lowest-order
contribution in $\alpha_s$ is singular at low momenta. In the
Factorization approach of refs.\,\cite{BBNS1,BBNS2} the amplitudes
for these processes are considered to be uncalculable. In the pQCD
approach~\cite{pqcd} Sudakov effects are invoked to regulate these
singularities and we have criticized the reliability of this
procedure in ref.\,\cite{dgs} (for a study of similar problems in
higher-twist contributions to the pion's electromagnetic form
factor, see ref.~\cite{pionff}). For the $\bwg$ decay the
situation is different, the lowest order contribution has no
singularity and there is no enhancement from non-perturbative
regions of phase-space.
 \label{item:resum}
\item What is a suitable choice of factorization scale $\mu_F$?

We argue in the following that it is (of order)
$\sqrt{m_b\lqcd}$.\label{item:muf}
\item Should the light-cone distribution amplitude be defined in QCD
or in the heavy-quark effective theory (HQET) ?

The choice is a pure matter of convenience, since the decay
amplitudes are independent of it. A different definition of the
light-cone distribution amplitude is compensated by a different
hard-scattering kernel, in order for their convolution to remain
identical in all cases. To connect our analysis with the results
from the SCET -- where heavy quarks are treated as in HQET -- we
find it convenient to define the distribution amplitude in the
heavy-quark effective theory. In addition, since the factorization
scale is much lower than $m_b$, it is natural to treat the
light-cone wave function in the framework of the HQET.
\item Are the one-loop results large? Does the resummation
mentioned in item~\ref{item:resum} make a large difference?

We find that the one-loop corrections are typically of the order
of a few times 10\% and are sensitive to the choice of
distribution amplitude. In particular, the corrections depend
sensitively on the two parameters  $\lambda_B^{(1)}$ and
$\lambda_B^{(2)}$ defined in eq.\,(\ref{eq:lambdaBn}). Resummation
of the large logarithms typically changes the results by up to
30\% or so of the one-loop correction.
\end{enumerate}
In this study, we neglect terms which are suppressed by a power of
$m_b$. In a number of the important two-body non-leptonic decay
channels the higher-twist terms are ``chirally enhanced" or have
larger CKM-matrix elements and so can give significant
contributions. The mass singularities in general do not factorize
for the higher-twist terms, so that the Factorization framework
described above has to be extended (for recent progress see
ref.\,\cite{higher}). Here we focus on the higher-order
perturbative corrections at leading twist.

Some of the above questions have been investigated in previous
studies, from which we have benefited considerably. Korchemsky,
Pirjol and Yan performed a detailed study of this decay process in
ref.~\cite{KPY} and concluded that a consistent factorization
formula can only be derived at one-loop order if the
hard-scattering kernel and the $B$-meson distribution amplitude
include a dependence on the transverse momenta. We disagree with
this conclusion. In sec.\,\ref{sec:th1loop} we perform the
matching explicitly, with the distribution amplitude defined as in
refs.~\cite{BBNS1,BF,GN}~\footnote{At leading twist, two
distribution amplitudes can actually be defined from
eq.~(\ref{eq:Bmatrixelem}) for the $B$-meson (see
eq.\,(\ref{eq:phi+-def})\,). However, only one contributes to the
$\bwg$ decay, and we shall call it the $B$-meson distribution
amplitude (or light-cone wave function) in the remainder of this
paper.} from the non-local matrix element:
\begin{equation} \label{eq:Bmatrixelem}
\Phi^B_{\alpha\beta}(\tp)= \int dz_- e^{i\tp z_-}
  \left.\langle 0| \bar{u}_\beta(z) [z,0] b_\alpha(0) | B
  \rangle\right|_{z_+,z_\perp=0}\,,
\end{equation}
where $[z,0]$ denotes a path-ordered exponential. (Here and for
the remainder of the paper we find it convenient to write the
label $B$ on the distribution amplitude as a superscript rather
than a subscript.) We find that it is consistent for the wave
function ($\Phi^B$) to depend on a single component of momentum
$\tilde k_+$, and that the amplitude of the decay process can be
expressed as a convolution over the single variable $\tp$:
\begin{equation}
F^{B}_\mu =\int \frac{d\tp}{2\pi} \Phi^B_{\alpha\beta}(\tp)
T_{\beta\alpha}(\tp)\,.
\end{equation}
Of course $\Phi^B$ does depend on the details of the internal
dynamics of the $B$-meson in a complicated way, which includes a
dependence on the transverse momentum of its constituents. This
does not matter however ($\Phi^B$ is a quantity which must be
determined non-perturbatively in any case). The key point is that
the hard-scattering kernel only depends on $\tilde k_+$. We stress
that, in general, $\tilde k_+$ should  not be identified with a
component of the momentum of any particular constituent of the
$B$-meson ($\tilde k_+$ is defined through
eq.~(\ref{eq:Bmatrixelem})\,).

The remainder of the paper is organized as follows. In the next
section we briefly discuss the kinematics for the decay $\bwg$. We
perform the matching, establish factorization and calculate the
hard-scattering amplitudes at tree level and at one-loop order in
secs.\,\ref{sec:tree} and \ref{sec:th1loop} respectively. In
sec.\,\ref{sec:resum} we use the SCET to resum the large
logarithms in the hard-scattering amplitude. We perform a brief
phenomenological study in sec.\,\ref{sec:phenom} and in
sec.~\ref{sec:concs} we present our conclusions and discuss open
questions.

\section{Kinematics of the \boldmath{$\bwg$}
Decay}\label{sec:kinematics}

\begin{figure}[t]
\begin{center}
\includegraphics[height=5cm]{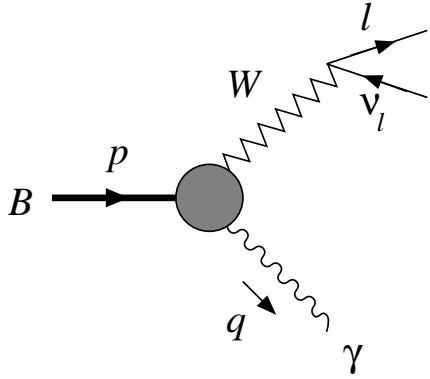}
\caption{Schematic representation of the $\bwg$ decay. The
momentum of the $B$-meson is denoted by $p$ and that of the photon
($\gamma$) by $q$.\label{fig:bwg}}
\end{center}
\end{figure}

The $\bwg$ decay is illustrated schematically in
fig.~\ref{fig:bwg}.  The momentum of the $B$-meson is denoted by
$p^\mu=M_B v^\mu$, where the four-velocity $v$ satisfies $v^2=1$.
In the following, unless otherwise stated, we will work in the
rest-frame of the $B$-meson, so that $v=(1,\vec 0\,)$. The
momentum and polarization vector of the photon are denoted by $q$
and $\varepsilon^*$ respectively. The energy of the photon is
given by
\begin{equation}
E_\gamma=v \cdot q= \frac{M_B^2-(p-q)^2}{2M_B}\le\frac{M_B}{2}\ .
\end{equation}
We consider decays in which the energy of both the photon and the
lepton pair is large, of order $M_B$, and we neglect $m_l/M_B$,
where $m_l$ is the mass of the lepton.

The hadronic matrix element for the decay $\bwg$ can be written in
terms of two form factors $F_V$ and $F_A$:
\begin{eqnarray}\label{eq:fsdef}
\frac{1}{\sqrt{4\pi\alpha}}\langle \gamma(\varepsilon^\ast,q)|
\bar{u}\gamma_\mu(1-\gamma_5) b | \bar{B}(p)\rangle=\\ &&
\hspace{-1.5in}\epsilon_{\mu\nu\rho\sigma} \varepsilon^{*\nu}
v^\rho q^\sigma
     F_V(E_\gamma) + i [\varepsilon^*_\mu (v\cdot q) - q_\mu (v\cdot \varepsilon^*)]
     F_A(E_\gamma)\,.\nonumber
\end{eqnarray}

For light-cone dominated processes, such as the $\bwg$ decay being
studied in this paper, it is convenient to introduce the
light-cone coordinates $l=(l_+,l_-,\vec{l}_\perp)$, defined by
\begin{equation}
l_{\pm}=\frac{l_0 \pm l_3}{\sqrt{2}}\,, \qquad \pv{l}=(l_1,l_2)\,,
\qquad l^2= 2l_+l_- - \pv{l}^{\ 2}\,.
\end{equation}
In the rest-frame of the $B$-meson, we choose the photon's
momentum to be in the minus direction so that:
\begin{equation}
p=(M_B/\sqrt{2}, M_B/\sqrt{2}, \pv{0}) \qquad\textrm{and}\qquad
q=(0, q_-, \pv{0})\,.
\end{equation}

We define the light-cone distribution amplitude of a state $H$
which contains the $b$-quark by
\begin{equation}\label{eq:phihdef}
\Phi^H_{\alpha\beta}(\tp)= \int dz_-\ e^{i\tp z_-}
  \left.\langle 0| \bar{u}_\beta(z) [z,0] b_\alpha(0) | H
  \rangle\right|_{z_+,z_\perp=0}\,,
\end{equation}
where $u,b$ are the quark fields and $\alpha,\beta$ are spinor
labels. $[z,0]$ denotes the path-ordered exponential ${\cal
P}\exp[-ig_s\int_0^z dz^\mu A_\mu(z)]$ and $g_s$ is the strong
coupling constant. In the following we will use the terms
\textit{light-cone distribution amplitude} and \textit{light-cone
wave function} interchangeably. In both cases the terms denote
$\Phi^H_{\alpha\beta}(\tp)$ as defined in eq.\,(\ref{eq:phihdef}).

$F^H_\mu$ is defined to be the matrix element of the weak $b\to u$
current,
\begin{equation}
F^{H}_\mu \equiv \langle \gamma(\varepsilon^*,q) |
\bar{u}\gamma_{\mu\,L} b | H \rangle\ ,
\end{equation}
where $\gamma_{\mu\,L}\equiv\gamma_\mu(1-\gamma_5)$. The question
which we investigate in this paper is whether, up to one-loop
order in perturbation theory, the matrix element can be written in
the \textit{factorized} form
\begin{equation}
F^{H}_\mu =\int \frac{d\tp}{2\pi} \Phi^H_{\alpha\beta}(\tp)
T_{\beta\alpha}(\tp)\,, \label{eq:factorH}
\end{equation}
where the hard-scattering amplitude $T$ does not depend on the
external state ($H$) and is a function of hard scales only. We
also evaluate $T$ up to one-loop order in perturbation theory.

Of course we are actually interested in the case where $H$ is the
$B$-meson. However, in the perturbative evaluation of $T$ we
exploit the independence of $T$ on the external state and choose a
convenient partonic state $H$ (in most cases we will take $H$ to
consist of a $b$-quark and a $\bar u$-antiquark). It is for this
reason that we introduce the notation with the general superscript
$H$.

The matrix element on the right-hand side of
eq.~(\ref{eq:Bmatrixelem}) can be written in terms of two
light-cone distribution amplitudes $\tilde\Phi_+^B$ and
$\tilde\Phi_-^B$~\cite{GN} and following ref.~\cite{BF} we
write~\footnote{Since in most of this paper we work in momentum
space, we introduce a tilde on $M,T$ and $\Phi_\pm^B$ in
coordinate space, and remove the tilde in momentum space.}
\begin{eqnarray}
\tilde M(z)&=&\langle 0| \bar{u}_\beta(z) [z,0] b_\alpha(0) |
\bar{B}(p) \rangle\nonumber\\ & = & -i\frac{f_B M_B}{4}
  \left[\frac{1+\dirac{v}}{2}
  \left(2\tilde\Phi_+^B + \frac{\dirac{z}}{t}
     (\tilde\Phi_-^B - \tilde\Phi_+^B)\right)\gamma_5\right]_{\alpha\beta}
\label{eq:phi+-def}\end{eqnarray} where $t=v\cdot z$ and
$\tilde\Phi_+^B$ and $\tilde\Phi_-^B$ depend on the coordinates
$z$ (they can be obtained from $\Phi^{H=B}_{\alpha\beta}$ by
Fourier transforming and projecting over the Dirac structure). We
will convolute this matrix element with a hard-scattering
amplitude $\tilde T(z)$:
\begin{equation}
\int d^4z \, \tilde M(z) \tilde T(z)= \int \frac{d^4
\tilde{k}}{(2\pi)^4} T(\tilde{k}) \int d^4z
   e^{i\tilde{k}\cdot z} \tilde M(z)\,.
\end{equation}
If $T(\tilde k)$ depends on a single component ($\tilde k_+$ in
our case), we can integrate the remaining components, which
corresponds to setting $z_+$ and $z_\perp$ to 0 in the matrix
element $\tilde M$ (the corresponding process is therefore a
light-cone dominated one). We will see explicitly in the following
sections that, at leading twist, it is only the function
$\Phi_+^B(\tilde k_+)$ which contributes to the form factors $F_V$
and $F_A$, and that $F_V=F_A$.

It may be helpful to stress the distinction between $\tilde k_+$
and the kinematical variables of the initial state. The light-cone
wave function $\Phi_{\alpha\beta}^H$ depends on the latter in a
complicated and non-perturbative way. An important goal of our
investigation is whether it is sufficient to introduce a
light-cone wave function as in eq.~(\ref{eq:phihdef}) depending
only on the + component of $\tilde k$ (as well as on all
components of the momenta in $H$) or whether a generalization of
eq.~(\ref{eq:phihdef}) is necessary. We will see that, up to
one-loop order at least, eq.~(\ref{eq:phihdef}) is sufficient.

\section{Factorization at Tree Level}\label{sec:tree}

\begin{figure}[t]
\begin{center}
\includegraphics[height=5cm]{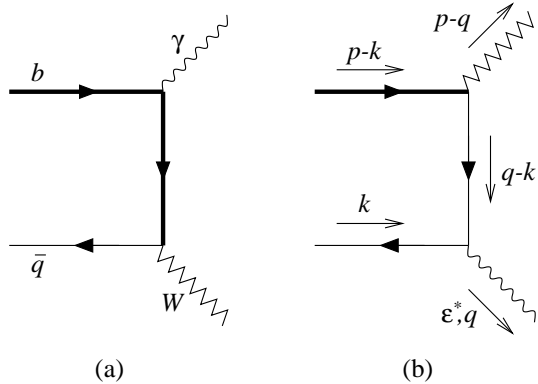}
\caption{Tree-level diagrams for the $\bwg$ decay with the
two-parton external state $|b\bar u\rangle$.} \label{fig:treediag}
\end{center}
\end{figure}

In this section we study factorization for the process $\bwg$ at
tree level. This is straightforward and the result is
well-known~\cite{KPY,DP}. However, the calculation is instructive
and so we exhibit the ingredients systematically. This will help
make our presentation of the one-loop calculation in
sec.~\ref{sec:th1loop} below clearer. Of course many features of
the generic calculation are absent at tree level. We do not
encounter mass singularities so that factorization at tree level
is guaranteed, and we do not have ultraviolet divergences and
hence do not have to choose a factorization scale. These questions
will arise at one-loop order in sec.~\ref{sec:th1loop}.

We wish to evaluate the hard-scattering kernel at tree level. For
the factorization formalism to be applicable, the hard-scattering
kernel must be independent of the infra-red effects contained in
the initial state and we are therefore free to choose this state
in any convenient manner. In this section we take the initial
state to be a light antiquark ($\bar u$) with momentum $k$ (with
the components of $k$ of order $\lqcd$) and a $b$-quark of
momentum $p-k$. The tree-level result for the hard-scattering
amplitude is given in eq.~(\ref{eq:Ttree}) below. It is also
instructive to see how the same hard-scattering amplitude arises
with a different initial state. In sec.~\ref{subsec:t0bug} we
illustrate this by obtaining the result of eq.~(\ref{eq:Ttree})
with a three-body, $b\bar ug$ (where $g$ represents a gluon),
initial state.

The two diagrams which contribute to the form factors at
tree-level are represented in Fig.~\ref{fig:treediag}. In
diagram~\ref{fig:treediag}(a) the (internal) $b$-quark propagator
is of $O(1/m_b)$ whereas in diagram~\ref{fig:treediag}(b) the
internal propagator, which is now the propagator of the $u$-quark,
is of $O(1/\lqcd)$. Thus diagram (a) is suppressed by one power of
the heavy-quark mass relative to diagram (b) and will be neglected
in the following. Also, in the evaluation of one-loop graphs in
sec.~\ref{sec:th1loop} below, we will only need to consider the
diagrams in which the photon is radiated from a light-quark.

In order to evaluate the hard-scattering amplitude we proceed in
the standard way:
\begin{enumerate}\vspace{-5pt}
\item[1.] compute the matrix element corresponding to the diagram of
fig~\ref{fig:treediag}(b);
\item[2.] evaluate the light-cone wave function at tree level;
\item[3.] combine the two to deduce the hard-scattering amplitude.
\end{enumerate}
We now carry out each of these three steps in turn.

\paragraph*{The Matrix Element at Tree-Level:}
At leading order in $1/m_b$, diagram~\ref{fig:treediag}(b) gives
the following contributions to the matrix element
\begin{eqnarray}
F_\mu^{(0)\,b\bar u}&\equiv& \la\gamma(\varepsilon^*,q)|\bar
u\gamma_{\mu\,L}b\,|b^S(p-k)\,\bar u^s(k)\rangle\nonumber\\
&=&-\frac{e_u}{2q_-k_+}\ \left\{\bar v^s(k)\,
\dirac\varepsilon^*\,\dirac{q}\,\gamma_{\mu\,L}\,u^S(p-k)\right\},
\label{eq:f0budef}\end{eqnarray} where $u$ and $v$ are the spinor
wave functions of the $b$ and $\bar u$ quarks respectively and $s$
and $S$ are spin labels. $e_u$ is the electric charge of a
$u$-quark.

\paragraph*{The Light-Cone Wave Function at Tree Level:} We define the
light-cone wave function, $\Phi^{b\bar u}(\tp)$, of the initial
state consisting of a $b$-quark and a $u$-antiquark as in
eq.~(\ref{eq:phihdef}):
\begin{equation}
\Phi^{b\bar u}_{\alpha\beta}(\tp)= \int dz_-\ e^{i\tp z_-}
  \left.\langle 0| \bar{u}_\beta(z) [z,0] b_\alpha(0) | b^S(p-k)\bar u^s(k)
  \rangle\right|_{z_+,z_\perp=0}\,.
\end{equation}
$\Phi^{b\bar u}$ depends on the initial state (for example in this
case it depends on $k$), but we leave this dependence implicit. At
tree level the wave function is readily found to be
\begin{equation}
\Phi^{(0)\,b\bar{u}}_{\alpha\beta}(\tp)=2\pi \delta(k_+-\tp)\
\bar{v}^s_\beta(k)\, u^S_\alpha(p-k)\,,
\label{eq:phi0budef}\end{equation} where the superscript $(0)$
denotes tree level.

\paragraph*{Matching and the Determination of the Hard-Scattering
Amplitude:} Writing the matrix element in the factorized form:
\begin{equation}
F^{(0)\,b\bar{u}}_\mu
  =\int \frac{d\tp}{2\pi} \Phi^{(0)\,b\bar{u}}_{\alpha\beta}(\tp)\,
  T^{(0)}_{\beta\alpha}(\tp)\,,
  \end{equation}
and taking the expressions for $F^{(0)\,b\bar u}$ and
$\Phi^{(0)\,b\bar{u}}$ from eqs.~(\ref{eq:f0budef}) and
(\ref{eq:phi0budef}) respectively, we obtain the hard-scattering
amplitude at tree level,
\begin{equation}
T^{(0)}_{\beta\alpha}(\tp)=-\frac{e_u}{2q_-\tp}
    [\dirac{\varepsilon}^* \dirac{q}
    \gamma_{\mu\,L}]_{\beta\alpha}\,.
\label{eq:Ttree}\end{equation} As expected the hard-scattering
amplitude depends only on $\tp$ (through the \textit{hard}
variable $q\cdot\tilde k$).

\subsection{Determining $T^{(0)}_{\beta\alpha}(\tp)$ with a
3-Body Initial State}\label{subsec:t0bug}

\begin{figure}[t]
\begin{center}
\includegraphics[height=5cm]{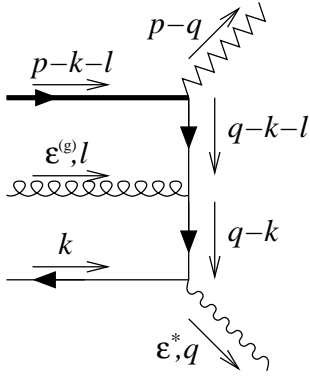}
\caption{Tree-level diagram for the $\bwg$ decay with the
three-parton external state $|b\bar ug\rangle$. Only the leading
twist contribution is shown.} \label{fig:threepartdiag}
\end{center}
\end{figure}

For the factorization formalism to be valid the hard-scattering
amplitude has to be independent of the external state. For example
it should be independent of the choice of external momentum $k$
(which we verify at one-loop order in sec.~\ref{sec:th1loop}).  In
this section we verify the independence of $T^{(0)}$ from the
external state in an instructive example. We take a three-body
initial state (see fig.~\ref{fig:threepartdiag}) consisting of a
$\bar u$-quark with momentum $k$, a gluon with momentum $l$ and a
$b$-quark with momentum $p-k-l$. Now the matrix element at
tree-level is
\begin{eqnarray}
F_\mu^{(0)\,b\bar ug}&\equiv& \la\gamma(\varepsilon^*,q)|\bar
u\gamma_{\mu\,L}b\,|b(p-k-l)\bar
u(k)g(\varepsilon^{(g)},l)\rangle\nonumber\\ &=& \frac{e_u g_s
\varepsilon^{(g)}_+(l)}{2q_-} \frac{1}{k_+(k+l)_+}
   \ \bar{v}^s(k)\dirac{\varepsilon}^* \dirac{q} \gamma_{\mu\,L}
   u^S(p-k-l)\,,
\end{eqnarray}
where $g_s$ is the strong coupling constant and
$\varepsilon^{(g)}$ is the polarisation vector of the gluon.

The lowest-order contribution to the light-cone wave function of
the three-body initial state is of $O(g_s)$ and corresponds to the
external gluon being annihilated by a gluon field present in the
path-ordered exponential $[z,0]$. The tree-level term in the wave
function is
\begin{eqnarray}
\Phi^{(0)\,b\bar{u}g}_{\alpha\beta}(\tp)
 &=&-ig_s \bar{v}^s_\beta(k) u^S_\alpha(p-k-l)
 \varepsilon^{(g)}_+(l)\nonumber\\
&&\hspace{0.4in}\times  \int dz_-\! \int_0^1 d\alpha\, z_-
e^{-i(k_++\alpha l_+ - \tp) z_-}
      \label{eq:alphaint}\\
 &&\hspace{-1.3in}= \frac{g_s}{l_+}\,(2\pi)[\delta(k_++l_+-\tp)-\delta(k_+-\tp)]
   \ \varepsilon^{(g)}_+(l)\, \bar{v}^s_\beta(k)\, u^S_\alpha(p-k-l)\,.
\end{eqnarray}
The integral over $\alpha$ in eq.\,(\ref{eq:alphaint}) is over the
position of the gluon field which annihilates the incoming gluon
(this position is taken to be $\alpha z$).

It is now straightforward to verify that
\begin{equation}
F^{(0)\,b\bar{u}g}_\mu
  =\int \frac{d\tp}{2\pi} \,\Phi^{(0)\,b\bar{u}g}_{\alpha\beta}(\tp)
  \,T^{(0)}_{\beta\alpha}(\tp)\,
\end{equation}
where $T^{(0)}_{\beta\alpha}(\tp)$ is given in
eq.~(\ref{eq:Ttree}). Thus we have verified that, as required, the
same hard scattering amplitude is obtained also with the
three-particle initial state.

\subsection{$\bwg$ at Tree Level}

Now that we have calculated the hard-scattering amplitude at
tree level, we can express the form-factors defined in
eq.~(\ref{eq:fsdef}) in terms of the light-cone wave function of
the $B$-meson. In this way we obtain the standard result:
\begin{equation}
F_A=F_V= \frac{f_B M_B\, Q_u}{2\sqrt{2}\,E_\gamma} \int_0^\infty
d\tp \frac{\Phi_+^B(\tp)}{\tp}+O(\as,1/M_B),
\end{equation}
where $Q_u=-2/3$ is the charge of the $\bar u$ antiquark in units
of the proton's charge. At this order the two form-factors are
equal and only depend on the first inverse moment of
$\Phi_+^B(\tilde k_+)$. This is similar to the corresponding
calculations of amplitudes of two-body non-leptonic $B$-decays.
The leading contribution to the term on the second line of
eq.\,(\ref{fff}) also depends on the first inverse moment of
$\Phi^B_+$.

\section{Factorization at One-Loop Order}
\label{sec:th1loop}

In this section we establish factorization for $\bwg$ decays at
one loop in perturbation theory and calculate the hard-scattering
amplitude at this order. We expand the matrix element, wave
function and hard-scattering amplitude in perturbation theory, so
that the factorization formula takes the schematic form,
\begin{eqnarray} \label{eq:expanFH}
F_\mu^H&=&F_\mu^{(0)\,H}+ F_\mu^{(1)\,H} + \cdots = \Phi^H \otimes
T \\ &=& [\Phi^{(0)\,H} \otimes T^{(0)}]
  + [\Phi^{(0)\,H} \otimes T^{(1)} + \Phi^{(1)\,H} \otimes T^{(0)} ] + \ldots
\nonumber\end{eqnarray} where $\otimes$ denotes the convolution,
and the superscripts indicate the power of $\as$. The
hard-scattering kernels $T^{(n)}$ contain only hard scales,
whereas the distribution amplitudes $\Phi^{(n)}$ absorb all the
soft effects.

At tree level we did not have to specify precisely the criterion
used to separate ``soft" and ``hard" scales. As discussed in the
introduction, $\bwg$ is a three scale process, with a large scale
($m_b$), a small scale ($\lqcd$) and an intermediate scale
($\sqrt{2q\cdot\tilde k}$ where $\tilde k=(\tilde k_+,0,\vec
0_\perp)$). We assume that large values of $\tilde k_+$ are damped
by the hadronic dynamics so that $\tilde k_+$ is predominantly of
$O(\lqcd)$ and the convolution is dominated by the region in which
$q\cdot\tilde k=O(m_b\lqcd)$, and that $m_b\lqcd$ is a
sufficiently large scale for perturbation theory to be applicable.
We will argue below that it is convenient to choose the
factorization scale $\mu_f$ to be of $O(\sqrt{m_b\lqcd})$. In this
section we evaluate $T^{(1)}$ and demonstrate that it is free of
any infrared effects. $T^{(1)}$ does however, contain large
logarithms which will have to be resummed. This will be discussed
in sec.~\ref{sec:resum}.

As in sec.~\ref{sec:tree}, we evaluate $T^{(1)}$ by taking
$H=|b^S(p-k)\,\bar u^s(k)\rangle$. The components of $k$ are taken
to be of $O(\lqcd)$, but the result for $T^{(1)}$ does not depend
on the precise choice. $T^{(1)}$ is obtained by using
eq.~(\ref{eq:expanFH})
\begin{equation}
\Phi^{(0)\,H} \otimes T^{(1)}=F^{(1)\,H}_\mu - \Phi^{(1)\,H}
\otimes T^{(0)}, \label{eq:T1matching}\end{equation} so that, at
one-loop order we need to evaluate both $\Phi^{(1)\,H}$ and
$F^{(1)\,H}_\mu$. There are mass singularities present in both the
terms on the right-hand side of eq.~(\ref{eq:T1matching}), but we
shall demonstrate that they cancel in the difference. Indeed, as
will become clearer below, it is convenient to consider the
difference in eq.~(\ref{eq:T1matching}) diagram by
diagram~\footnote{We perform the calculations in the Feynman
Gauge.}. The mass singularities cancel diagram by diagram, and it
is of course necessary to regulate them in the same way in both
$F^{(1)\,H}_\mu$ and $\Phi^{(1)\,H} \otimes T^{(0)}$. Unless
explicitly stated to the contrary, we regulate the collinear
divergences by giving the $\bar u$ antiquark a mass ($k^2=m^2$)
and the infrared divergences by giving the gluon a mass $\lambda$.
The hard-scattering amplitude is independent of $m$ and $\lambda$.
The ultraviolet divergences are regulated by dimensional
regularization (we work in $d=4-\varepsilon$ dimensions) and we
use the $\overline{\textrm{MS}}$ renormalization scheme, by
redefining $\mu^2\to\mu^2e^{\gamma_E}/4\pi$, where $\gamma_E$ is
Euler-Mascheroni constant, and subtracting divergences
proportional to powers of $N^{UV}_\varepsilon=2/\varepsilon$. (At
one loop, if only single poles are present, this involves the
subtraction of terms of the form
$2/\varepsilon-\gamma_E+\log(4\pi)$.)

Eq.~(\ref{eq:T1matching}), together with the tree-level
expressions for the light-cone wave function for the $b\bar u$
initial state in eq.~(\ref{eq:phi0budef}) and the hard-scattering
amplitude in eq.~(\ref{eq:Ttree}), gives the following expression
for $T^{(1)}$
\begin{equation}\bar{v}^s(k) T^{(1)} u^S(p-k)=
F^{(1)\,b\bar u}_\mu
  - \int \frac{d\tp}{2\pi} \Phi^{(1)\,b\bar{u}}_{\alpha\beta}(\tp)
  \frac{(-\,e_u)}{2q_-\tp}
        [\dirac{\varepsilon}^* \dirac{q}
        \gamma_{\mu\,L}]_{\beta\alpha}\ .
\label{eq:T1expr}\end{equation}

We now evaluate the contributions to each of the two terms on the
right-hand side of eq.~(\ref{eq:T1expr}) from each of the
diagrams. Since all the calculations are performed with an
external state of a $b$ quark and a $\bar u$ antiquark, for
compactness of notation we will omit the $b\bar u$ label on $F$
and $\Phi$ in the remainder of this section.

\subsection{Electromagnetic Vertex}
\label{subsec:emvertex}

\begin{figure}[t]
\begin{center}
\includegraphics[height=5cm]{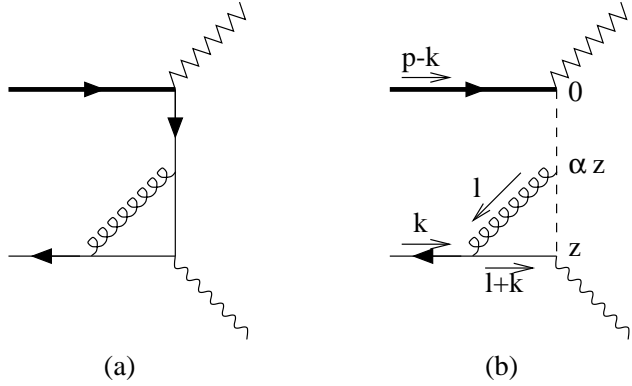}
\caption{Electromagnetic vertex: contributions to the form factors
(left) and $\Phi^{(1)\,\textrm{em}}\otimes T^{(0)}$ (right). The
dashed line in the right-hand diagram represents the path-ordered
exponential.} \label{fig:emvertex}
\end{center}
\end{figure}

The contribution to $F^{(1)}_\mu$ from the vertex correction to
the electromagnetic vertex (the diagram in
fig.~\ref{fig:emvertex}(a)\,) is
\begin{eqnarray}
F_\mu^{(1)\,{\rm em}}&=& -\frac{ig_s^2C_Fe_u}{2q_-k_+}
\int\frac{d^dl}{(2\pi)^d}\times \nonumber\\
&&\hspace{0in}\frac{\bar
v^s(k)\gamma^\rho(\dirac{k}+\dirac{l})\dirac{\varepsilon}^\ast(\dirac{q}-\dirac{k}-\dirac{l})
\gamma_\rho\dirac{q}\gamma_{\mu\,L}u^S(p-k)}{(l^2+i\varepsilon)((k+l)^2-m^2+i\varepsilon)
((q-k-l)^2-m^2+i\varepsilon)}\,, \label{eq:emint}\end{eqnarray}
where we have neglected terms which are suppressed by $\lqcd/m_b$.
Evaluating the integral, we obtain
\begin{equation}
F_\mu^{(1)\,{\rm em}}=  \frac{\as C_F}{4\pi}
\frac{(-\,e_u)}{2q_-k_+}
   \{\bar{v}^s \dirac{\varepsilon}^* \dirac{q} \gamma_{\mu\,L}
   u^S\}
   \left[N_\varepsilon^{UV} - \log\frac{2k\cdot q}{\mu_R^2}+2\log\frac{2k\cdot
   q}{m^2}\right]\,,
\label{eq:f1emresult}\end{equation} where $\mu_R$ is the
renormalization scale (since the electromagnetic current does not
get renormalized, the dependence on $\mu_R$ cancels with the
corresponding contributions from the wave function
renormalization)~\footnote{In table~1. of ref.\cite{KPY} there is
an additional $-1$ in the square parentheses. We have traced this
discrepancy to the expression for $J_1^{(b)}$ in eq.\,(A9) of
ref.\,\cite{KPY}.}. If we take the small mass limit by decreasing
$k_-$ and $k_\perp$ while keeping $k_+$ fixed (and of
$O(\lqcd)$\,), then the mass singularities of the form $\log(m^2)$
in eq.~(\ref{eq:f1emresult}) come from the collinear region of
phase-space, $l_+=O(\lambda^0),\ l_-=O(\lambda^2)$ and
$l_\perp=O(\lambda)$ with $\lambda\to 0$.

The corresponding contribution to the distribution amplitude is:
\begin{eqnarray}
\Phi^{(1)\,{\rm em}}_{\alpha\beta}(\tp)
  &=&C_F g_s^2 \int dz_- e^{iz_-\tp}
    \int_0^1 d\alpha z_- \\
  &&\quad \int \frac{d^dl}{(2\pi)^d}
      e^{-iz_-(k+l-\alpha\,l)_+}
      \left[\bar{v}^s\gamma_+\frac{1}{\dirac{l}+\dirac{k}+m}\right]_\beta
      u^S_\alpha \frac{1}{l^2}
      \nonumber\\
  &=&-iC_F g_s^2 \int \frac{d^dl}{(2\pi)^d} \frac{1}{l_+}
    (2\pi)[\delta(k_+-\tp)-\delta(l_++k_+-\tp)] \nonumber\\
&&\qquad  \times [\bar{v}^s\gamma_+(\dirac{l}+\dirac{k})]_\beta
u^S_\alpha
       \frac{1}{[(k+l)^2-m^2][l^2]}\,.\label{eq:emmatch}
\end{eqnarray}
$\alpha\,z_-$ is the position of the gluon field in the
path-ordered exponential (drawn as a dashed line in
Fig.~\ref{fig:emvertex}(b)) used to construct the gluon
propagator. After the integration over $\alpha$, followed by that
over $z_-$, we obtain eq.~(\ref{eq:emmatch}). This equation
underlines again that $\tilde k_+$ cannot be identified with
$k_+$, and is not a priori of $O(\lqcd)$ (it is $k_+$ which is
taken to be of $O(\lqcd)$). However, the convolution with
$T^{(0)}$ (which is proportional to $1/\tp$) damps large values of
$\tp$ and is indeed dominated by the region in which $\tilde k_+$
is of $O(\lqcd)$. This is also true for the other one-loop
diagrams, as can be seen from the expressions given in the
following sections.

Using the expression for $\Phi^{(1)\,{\textrm{em}}}$ in
eq.~(\ref{eq:emmatch}), the contribution to $\Phi^{(1)}\otimes
T^{(0)}$ is found to be:
\begin{eqnarray}
\Phi^{(1)\,{\rm em}}\otimes T^{(0)}&=&-
\frac{ig_s^2C_Fe_u}{2q_-k_+} \int\frac{d^dl}{(2\pi)^d}
\times\nonumber\\ &&\hspace{-0.1in}\frac{\bar
v^s(k)\gamma^\rho(\dirac{k}+\dirac{l})\dirac{\varepsilon}^\ast\dirac{q}
\gamma_\rho\dirac{q}\gamma_{\mu\,L}u^S(p-k)}{(l^2+i\varepsilon)((k+l)^2-m^2+i\varepsilon)
(-2q_-(k_++l_+))}\,. \label{eq:emmatching}\end{eqnarray} Comparing
eqs.~(\ref{eq:emint}) and (\ref{eq:emmatching}) we find that, as
expected, they are related by the replacement of the internal
light-quark propagator by the eikonal approximation:
\begin{equation}
\frac{i(\dirac{q}-\dirac{k}-\dirac{l})}{(q-k-l)^2-m^2}\to
\frac{i\dirac{q}}{-2q_-(k_++l_+)}\,.
\label{eq:eikonal}\end{equation} The diagram in
fig.\,\ref{fig:emvertex}(b) represents
$\Phi^{(1)\,\textrm{em}}\otimes T^{(0)}$, with the dashed lines
representing the eikonal propagators.

In the collinear region of phase space, the replacement in
eq.\,(\ref{eq:eikonal}) could be made also in $F^{(1)\,{\rm
em}}_\mu$ and therefore we would expect that the collinear
singularities are the same in $F^{(1)\,{\rm em}}_\mu$ and
$\Phi^{(1)\,{\rm em}}\otimes T^{(0)}$, and this is indeed the
case. Evaluating the integral in eq.~(\ref{eq:emmatching}) we
obtain the result
\begin{equation}
\Phi^{(1)\,{\rm em}} \otimes T^{(0)}
 = -\frac{\as C_F}{4\pi} \frac{e_u}{2q_-k_+}
   \bar{v}^s \dirac{\varepsilon}^* \dirac{q} \gamma_{\mu\,L}u^S
   \left[2N_\varepsilon^{UV} - 2\log\frac{m^2}{\mu_F^2}+4\right]\,.
\end{equation} The integral has ultraviolet divergences which again we
regulate using dimensional regularization and renormalize using
the $\overline{\textrm{MS}}$ scheme with a renormalization scale
$\mu_F$. $\mu_F$ is the natural choice for the factorization
scale.

We see that, as anticipated, the mass singularities in
$F^{(1)\,{\rm em}}_\mu$ and $\Phi^{(1)\,{\rm em}} \otimes T^{(0)}$
are the same, and from their difference we obtain the following
contribution to the hard-scattering kernel:
\begin{equation}\label{eq:T1em}
\begin{array}{|c|}
\hline \displaystyle \rule[-0.25in]{0in}{0.6in}T^{(1)\,{\rm
em}}(\tp;\mu_F)=\frac{\as C_F}{4\pi}T^{(0)}(\tp)
  \left[2\log \frac{2\tilde{k}\cdot q}{\mu^2_F}
  - \log \frac{2\tilde{k}\cdot q}{\mu^2_R} - 4 \right]\ .\\
\hline
\end{array}
\end{equation}

\subsection{Wave Function Renormalization}

\begin{figure}[t]
\begin{center}
\includegraphics[height=5cm]{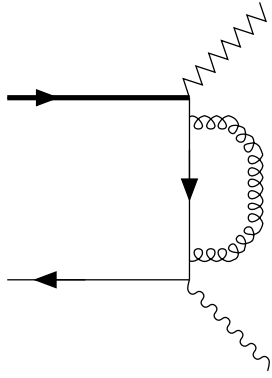}
\caption{Wave-function renormalization diagram for the internal light-quark line.}
\label{fig:selfdiag}
\end{center}
\end{figure}

The contribution of the diagram in Fig.~\ref{fig:selfdiag} to
$F^{(1)}_\mu$ can be readily evaluated:
\begin{equation}\label{eq:fse}
F^{(1)\,{\rm wfc}}_\mu=-\frac{\as C_F}{4\pi} \frac{e_u}{2q_-k_+}
   \bar{v}^s \dirac{\varepsilon}^* \dirac{q} \gamma_{\mu,L} u^S
   \left[-N_\varepsilon^{UV} +\log\frac{2k\cdot
   q}{\mu_R^2}-1\right]\ .
\end{equation}

There is no corresponding contribution to $\Phi^{(1)}$. The gluon
propagator in this case is $\langle 0|A_+(\alpha z_-) A_+(\beta
z_-)|0\rangle$, where $\alpha z_-$ and $\beta z_-$ are two points
on the line between the origin and $z$ ($\alpha,\beta$ are
integrated between 0 and 1, respecting the path-ordering). In the
Feynman gauge this propagator vanishes.

The contribution of Fig.~\ref{fig:selfdiag} to the hard-scattering
kernel is therefore:
\begin{equation}\label{eq:T1se}
\begin{array}{|c|}
\hline \displaystyle \rule[-0.25in]{0in}{0.6in}
T^{(1)\,{\rm wfc}}(\tp)=\frac{\as C_F}{4\pi}T^{(0)}(\tp)
  \left[\log \frac{2\tilde{k}\cdot q}{\mu^2_R} - 1 \right]\ ,\\
\hline
\end{array}
\end{equation}
which is free of mass singularities.

We now turn to the wave function renormalization of the external
quark fields. We start by considering the wave function
renormalization of the $\bar u$ antiquark,
\begin{eqnarray}\label{eq:fuwf}
F^{(1)\,\bar u{\rm wf}}_\mu&=&\left\{\frac{1}{2}[Z_2^{\bar u
}-1](\mu_R)\right\} F^{(0)}_\mu\,,\\ \Phi^{(1)\,\bar u{\rm wf}}
\otimes T^{(0)}
  &=& \left\{\frac{1}{2}[Z_2^{\bar u}-1](\mu_F)\right\}\Phi^{(0)}\otimes
  T^{(0)}\,.\label{eq:phiuwf}
\end{eqnarray}
The wave function renormalization constant $Z_2$ for a quark with
momentum $p$ and mass $m$ in QCD is defined here in terms of the
one-particle irreducible graphs $\Sigma$ by
\begin{equation}
Z_2=1+i\left.\frac{d\Sigma}{d\dirac{p}}\right|_{\dirac{p}=m}\,.
\label{eq:z2defq}\end{equation} Since infrared effects cancel in
the difference of the terms in eqs.~(\ref{eq:fuwf}) and
(\ref{eq:phiuwf}), it is convenient in this case to regulate the
infrared divergences by choosing the $\bar u$ antiquark to be
off-shell, $k^2\neq m^2$. Then
\begin{equation}
\frac{1}{2}[Z_2^{\bar u}-1](\mu)
  =  \frac{\as C_F}{4\pi}
  \left[-\frac{1}{2}N_\varepsilon^{UV}-2\log\left(1-\frac{k^2}{m^2}\right)
         +\log\frac{m}{\mu}-2
  \right]\,.
\label{eq:z2ubar}\end{equation} The corresponding contribution to
the hard-scattering kernel is
\begin{equation}
\begin{array}{|c|}
\hline \displaystyle\rule[-0.25in]{0in}{0.6in} T^{(1)\,\bar u{\rm
wf}}(\tp;\mu_F) =\frac{\as C_F}{4\pi}T^{(0)}(\tp)
  \frac{1}{2}\log \frac{\mu^2_F}{\mu^2_R}\ .\\
\hline
\end{array}
\label{eq:T1ubarwf}
\end{equation}

For the renormalization of the external $b$-quark field, the
contribution to $\Phi^{(1)\,Q{\rm wf}} \otimes T^{(0)}$
corresponds to the HQET. The superscript $Q$ denotes
\textit{heavy-quark}, as defined in the HQET, and is used to
distinguish it from the $b$-quark in QCD. We have
\begin{eqnarray}\label{eq:fbwf}
F^{(1)\,b{\rm wf}}&=& \left\{\frac{1}{2}[Z_2^{b}-1](\mu_R)
  \right\} F^{(0)}\,,\\
\Phi^{(1)\,Q{\rm wf}} \otimes T^{(0)}
  &=& \left\{\frac{1}{2}[Z_2^{Q}-1](\mu_F)
  \right\}\Phi^{(0)}\otimes T^{(0)}\,,
\label{eq:phiQ}\end{eqnarray} where $Z_2^b$ is given by
eq.~(\ref{eq:z2ubar}) which we now rewrite in the form (we take
$p^2=m_b^2$)
\begin{equation}
\frac{1}{2}[Z_2^{b}-1](\mu_R)
  =  \frac{\as C_F}{4\pi}
  \left[-\frac{1}{2}N_\varepsilon^{UV}
   -2\log\frac{2\,v\cdot k}{\mu_R}
   +3\log\frac{m_b}{\mu_R}-2\right],
   \end{equation}
For a quark $Q$ with residual momentum $k$ in the HQET we define
the wave function renormalization constant through
\begin{equation}
Z_2^Q=1+i\left.\frac{d\Sigma}{d (v\cdot k)}\right|_{v\cdot k=0}
\label{eq:z2defQ}\end{equation} and find
\begin{equation}
\frac{1}{2}[Z_2^{Q}-1](\mu_F)
  =  \frac{\as C_F}{4\pi}
  \left[N_\varepsilon^{UV}-2\log\frac{2\,v\cdot k}{\mu_F}
   \right]\,.
\end{equation}

After renormalization, the contribution to the hard-scattering
kernel is
\begin{equation}
\begin{array}{|c|}
\hline \displaystyle\rule[-0.25in]{0in}{0.6in}
T^{(1)\,b{\rm wf}}(\tp;\mu_F)=\frac{\as C_F}{4\pi}T^{(0)}(\tp)\left\{
\log\frac{\mu_R^2}{\mu_F^2}+\frac32\log\frac{m_b^2}{\mu_R^2}-2\right\}\,.\\
\hline
\end{array}
\label{eq:T1bwf}\end{equation}

We end this subsection by using dimensional regularization to
regulate the infrared divergences, instead of taking the external
quarks to be off-shell. This provides further verification of the
independence of the hard-scattering amplitude from the
regularization, and allows us to compare our results with earlier
studies. Using dimensional regularization in QCD at the scale
$\mu$ we find:
\begin{equation}\label{eq:z2qcddr}
\frac12[Z_2^q-1](\mu)=\frac{\as C_F}{4\pi}\left[-\frac12
N_\varepsilon^{UV}+\frac{3}{2}\log\frac{m_q^2}{\mu^2}-N_\varepsilon^{IR}-2\right]
\end{equation}
for a quark $q$ of mass $m_q$ ($q=\bar u$ or $b$) with
$N_\varepsilon^{IR}=2/\varepsilon$ (recall that a rescaling of
$\mu$ has been implicitly performed and cancelled the contribution
proportional to $-\gamma_E+\log(4\pi)$). The result in
eq.\,(\ref{eq:z2qcddr}) agrees with eq.\,(29) of ref.\cite{SCET1},
but disagrees with that in table 1 of ref.\cite{KPY}. For the
heavy quark effective theory we find
\begin{equation}\label{eq:z2hqetdr}
\frac12[Z_2^Q-1](\mu)=\frac{\as C_F}{4\pi}\left[
N_\varepsilon^{UV}-N_\varepsilon^{IR}\right]\,.
\end{equation}
To compute the contribution of the $b$-quark wave function
renormalization graph to $T_H$ in dimensional regularization, we
take the difference of eqs.~(\ref{eq:z2qcddr}) (with the quark
mass set to $m_b$) and (\ref{eq:z2hqetdr}). Comparing the result
with eq.~(\ref{eq:T1bwf}), which is the same contribution
evaluated in the off-shell regularization scheme, we see that it
is indeed the same for $\mu=\mu_F=\mu_R$~\footnote{In the case of
dimensional regularization, there is only one scale $\mu$ for both
ultraviolet and infrared divergences. We need therefore to set the
factorization and the renormalization scales to the same value in
order to compare the hard-scattering kernel with the result
obtained in another scheme.}. Repeating the analysis for the wave
function renormalization graph for the light quark, we find
vanishing contributions in both cases. Thus, as expected, we
obtain the same result for the hard-scattering kernel for both
methods of regulating the mass singularities.

\subsection{Weak vertex}

\begin{figure}[t]
\begin{center}
\includegraphics[height=5cm]{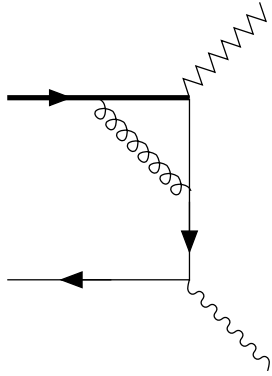}
\caption{Diagram representing the one-loop correction to the weak
vertex.} \label{fig:wkvertex}
\end{center}
\end{figure}

We now turn to the correction to the weak vertex illustrated in
the diagram in fig.~\ref{fig:wkvertex}. Evaluating this diagram we
find
\begin{eqnarray}\label{eq:fwk}
F^{(1)\,{\rm wk}}_\mu
  &=& \frac{\as C_F}{4\pi} F^{(0)}_\mu
   \Bigg[N_\varepsilon^{UV} - \log\frac{x}{\mu_R^2}
      -\frac{y}{x-y}\log\frac{y}{x}-\log^2\frac{y}{z}\\
&&\hspace{-1in}
      +2\log\frac{x}{y} \log\frac{y}{z}
      +2\log\frac{y}{z}+\log^2\frac{x}{y}
      +2{\rm Li}_2\left(1-\frac{y}{x}\right) + 4 {\rm Li}_2\left(1-\frac{x}{y}\right)
      -\pi^2
   \Bigg]+\ldots\nonumber
\end{eqnarray}
with
\begin{equation}\label{eq:xyzdef}
x=m_b^2\,, \qquad y=2m_b\eg\,\qquad\textrm{and} \qquad z=2(q\cdot
k)\,.
\end{equation}
The ellipses denote terms which are not relevant for the
discussion, since their Dirac structure $\Gamma$ is such that they
vanish when convoluted with the distribution amplitudes: ${\rm
Tr}[(1+\dirac{v})\gamma_5\Gamma]={\rm
Tr}[(1+\dirac{v})\gamma_+\gamma_5\Gamma]=0$.

In ref.~\cite{KPY} this diagram was evaluated in the HQET. The infrared
behaviour of eq.~(\ref{eq:fwk}):
\begin{equation} \label{eq:weakdblelogs}
-\log^2 k_+ + 2\log k_+ \log(\sqrt{2}\eg) - 2 \log k_+
\end{equation}
agrees with that in ref.~\cite{KPY}.

The corresponding contribution to $\Phi^{(1)}\otimes T^{(0)}$ is
\begin{eqnarray}\label{eq:phiwk}
&&\Phi^{(1)\,{\rm wk}}\otimes T^{(0)}
 = \frac{\as C_F}{4\pi}\,\left\{ -\frac{e_u}{2q_-k_+}
   \bar{v}^s \dirac{\varepsilon}^* \dirac{q} \gamma_{\mu\,L} u^S\right\} \\
&&\times
   \left[ -(N_\varepsilon^{UV})^2 -
   2N_\varepsilon^{UV}\log\frac{\mu_F}{\sqrt{2}k_+}-2\log^2\frac{\mu_F}{\sqrt{2}k_+}
   -\frac{3\pi^2}{4}
   \right]+\ldots\nonumber
\end{eqnarray}
We recall at this point that we are using the modified minimal
subtraction scheme ($\overline{\textrm{MS}}$). We have therefore
redefined $\mu^2\to\mu^2e^{\gamma_E}/4\pi$, and will subtract the
divergences proportional to powers of
$N_\varepsilon^{UV}=2/\varepsilon$.

The contribution to the hard-scattering kernel contains double
logarithms and is given by:
\begin{equation}
\begin{array}{|rcl|}
\hline \displaystyle\rule[-0.25in]{0in}{0.6in} T^{(1)\,{\rm wk}}
  &=&\displaystyle
\frac{\as C_F}{4\pi} T^{(0)}
   \Bigg[-
   \log\frac{m_b^2}{\mu_R^2}+\frac{1}{2}\log^2\frac{m_b^2}{\mu_F^2}\\
   &&\displaystyle\hspace{-0in}\rule[-0.25in]{0in}{0.6in}-
      \frac{\sqrt{2}q_-}{m_b-\sqrt{2}q_-}\log\frac{\sqrt{2}q_-}{m_b}
      +\log^2\frac{m_b}{\sqrt{2}\tp}\\
&& \displaystyle\rule[-0.25in]{0in}{0.6in}
      +2\log\frac{\mu_F^2}{\sqrt{2}q_-m_b} \log\frac{m_b}{\sqrt{2}\tp}
      +2\log\frac{m_b}{\sqrt{2}\tp}\\
&& \displaystyle \hspace{-0.6in}+\log^2\frac{m_b}{\sqrt{2}q_-}
      +2{\rm Li}_2\left(1-\frac{\sqrt{2}q_-}{m_b}\right) +
      4 {\rm Li}_2\left(1-\frac{m_b}{\sqrt{2}q_-}\right)
      -\frac{\pi^2}{4}\rule[-0.25in]{0in}{0.6in}
   \Bigg]\,.\\
\hline
\end{array}
\label{eq:T1wk}
\end{equation}

\subsection{Box diagram}

\begin{figure}[t]
\begin{center}
\includegraphics[height=5cm]{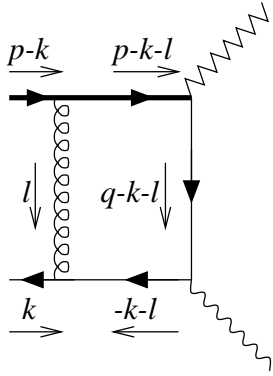}
\caption{Box diagram.\label{fig:box}}
\end{center}
\end{figure}

The final one-loop graph is the box diagram in fig.~\ref{fig:box}.
It has been argued in ref.~\cite{KPY} that the structure of this
graph requires the introduction of a transverse momentum
dependence into the $B$-meson wave function, and hence invalidates
the factorization formalism encapsulated in
eq.~(\ref{eq:factorH}). We disagree with this conclusion. We will
see that the leading-twist contribution from the box diagram is a
soft effect which is absorbed completely into the light-cone wave
function (i.e. the contributions to $F_\mu^{(1)}$ and
$\Phi^{(1)}\otimes T^{(0)}$ are the same). The question is not
whether the box diagram exhibits a dependence on the transverse
components of the external quarks (which in general it does), but
whether this dependence matches the one induced by the
distribution amplitude. This is indeed the case and therefore we
find no breakdown of the QCD factorization framework at one-loop
for $\bwg$ decays.

The contribution to $F_\mu^{(1)}$ from the box diagram in
fig.~\ref{fig:box} is
\begin{eqnarray}\label{eq:f1box}
F^{(1)\,\textrm{box}}_\mu&=&
  -ig_s^2C_Fe_u \int \frac{d^4l}{(2\pi)^4}\times\\
&&\hspace{-0.9in}\frac{
\bar{v}^s\gamma^\rho\,[-\dirac{l}-\dirac{k}+m]
\dirac{\varepsilon}^\ast\,[\dirac{q}-\dirac{k}-\dirac{l}+m]\gamma_{\mu\,L}
[\dirac{p}-\dirac{k}-\dirac{l}+m_b]\gamma_\rho u^S}
{[(l+k)^2-m^2+i\varepsilon][l^2+i\varepsilon]
[(q-k-l)^2-m^2+i\varepsilon][(p-k-l)^2-m_b^2+i\varepsilon]}\,.
\nonumber
\end{eqnarray}
We are interested in the leading twist contribution to
$F^{(1)}_\mu$, which is of $O(1/k_+)$. At one-loop level the box
diagram is the only one with no propagator of order $1/\lqcd$
outside of the loop. The $1/k_+$ enhancement is therefore more
difficult to obtain than for the other diagrams; it must come from
singular regions of phase space in the loop integral. We now
investigate what these regions are.

Consider the region of phase-space in which the components of $l$
are of order $k_+\sim k_-\sim \lqcd$. Power counting shows that
this region gives a leading twist contribution. However in this
region the $b$-quark propagator can be approximated by that of the
HQET and
\begin{equation}
\frac{\dirac{q}-\dirac{k}-\dirac{l}+m}{(q-k-l)^2-m^2+i\varepsilon}
\to \frac{\dirac{q}}{-2q_-(k_++l_+)}\ ,
\label{eq:boxeikonal}\end{equation}i.e. the light-quark propagator
joining the electromagnetic and weak vertices can be replaced by
its eikonal approximation. We therefore find that the contribution
from this soft region is equal to the corresponding term in
$\Phi^{(1)\,\textrm{box}}\otimes T^{(0)}$. Indeed we obtain at
leading twist
\begin{equation}
F^{(1)\,\textrm{box}}_\mu=\Phi^{(1)\,\textrm{box}}\otimes
T^{(0)}\,,
\end{equation}
so that the corresponding contribution to $T^{(1)}$ is zero,
\begin{equation}\label{eq:T1box}
\begin{array}{|c|}
\hline \displaystyle \rule[-0.in]{0in}{0.2in}
T^{(1)\,\textrm{box}}=0\,.
\\ \hline
\end{array}
\end{equation}

We conclude this subsection by briefly illustrating the
suppression of contributions from regions of phase-space which
would lead to a non-zero contribution to $T^{(1)\,\textrm{box}}$.
As an example consider the contribution from the collinear region
in which $l_-\sim O(m_b),\,l_+\sim O(\lqcd^2/m_b)$ and
$l_\perp\sim O(\lqcd)$. In this case the phase-space is of
$O(\lqcd^4)$ and the four factors in the denominator in
eq.~(\ref{eq:f1box}) are of order $\lqcd m_b ,\lqcd^2,\lqcd m_b$
and $m_b^2$ respectively. This region therefore does not give a
contribution of $O(1/\lqcd)$ (in addition there is a further
suppression factor of $\lqcd$ from the numerator). A detailed
study of $F^{(1)\,\textrm{box}}_\mu$ and
$\Phi^{(1)\,\textrm{box}}\otimes T^{(0)}$ shows that they are
equal at leading twist.

\subsection{The Hard-Scattering Kernel}

We obtain the complete one-loop hard-scattering kernel by summing
the contributions from the electromagnetic vertex,
eq.~(\ref{eq:T1em}), the wave function renormalization of the
collinear light quark, eq.~(\ref{eq:T1se}), the wave function
renormalization of the external $\bar u$-antiquark and the
$b$-quark, eqs.~(\ref{eq:T1ubarwf}) and (\ref{eq:T1bwf})
respectively, the weak vertex, eq.~(\ref{eq:T1wk}) and the box
diagram eq.~(\ref{eq:T1box}). We now convolute the result with the
wave function of the $B$-meson, using the decomposition in
eq.~(\ref{eq:phi+-def}), obtaining the leading twist one-loop
expression for the form factors:
\begin{equation}
F_A(\eg)=F_V(\eg)= \int d\tp \Phi_+^B(\tp;\mu_F)
T(\tp,\eg;\mu_F)\,,
\label{eq:favoneloop}\end{equation} where
\begin{equation}
T(\tp,\eg;\mu_F)=\frac{f_B M_B Q_u}{2\sqrt{2}\eg} \frac{1}{\tp}
  \left[1+\frac{\as C_F}{4\pi} K(\tp,\eg;\mu_F) \right]
\label{eq:aux}\end{equation} and
\begin{eqnarray}
K(\tp,\eg;\mu_F) &=&
      \log^2\frac{2\tp q_-}{\mu_F^2}
      -\frac{1}{2}\log^2\frac{m_b^2}{\mu_F^2}
      +\frac{5}{2}\log\frac{m_b^2}{\mu_F^2}
      +2\log\frac{m_b^2}{\mu_F^2}\log\frac{m_b}{\sqrt{2}q_-}\nonumber\\
&&    -2\log\frac{m_b}{\sqrt{2}q_-}
      -\frac{\sqrt{2}q_-}{m_b-\sqrt{2}q_-}\log\frac{\sqrt{2}q_-}{m_b}
\label{eq:t1oneloop}
\\
&&   +2{\rm Li}_2\left(1-\frac{\sqrt{2}q_-}{m_b}\right) + 4 {\rm
Li}_2\left(1-\frac{m_b}{\sqrt{2}q_-}\right)
      -\frac{\pi^2}{4}-7
   \nonumber\,.
\end{eqnarray}
The dependence on renormalization scale $\mu_R$ has canceled as it
must do.

The formulae in eqs.~(\ref{eq:favoneloop}) -- (\ref{eq:t1oneloop})
above give the one-loop factorized expression for the form factors
$F_A$ and $F_V$. The expression for $T$ in eqs.~(\ref{eq:aux}) and
(\ref{eq:t1oneloop}) differs from the corresponding result in
refs.~\cite{KPY,DP}. Of course $T$ depends on the definition of
the light-cone wave function and we present our definition in
sec.~\ref{sec:kinematics}. With this definition there is no need
to introduce a dependence on transverse momenta, at least at
one-loop order. We disagree also with the formula proposed in
eq.~(12) of ref.~\cite{DP}.

The large double and single logarithms present in
eq.\,(\ref{eq:t1oneloop}), $\log^2(\sqrt{2}\tp/m_b)$ and
$\log(\sqrt{2}\tp/m_b)$, need to be resummed and this is the
subject of the next section.

\section{Resummation of Large Logarithms}\label{sec:resum}

In order for the expressions for the hard-scattering amplitude to
be useful phenomenologically, the large logarithms need to be
resummed. This is true in particular for the Sudakov effects
associated with the $b\to u$ weak vertex. An elegant method for
performing this resummation exploits the effective field theory
describing the interaction of (infinitely) heavy quarks and
(massless) collinear quarks with soft and collinear
gluons~\cite{SCET2,SCET1}. Alternatively one could perform the
resummation by using the ``Wilson-Line"
formalism~\cite{KPY,GK,KS}. We now adapt the discussion of
ref.~\cite{SCET1} to the $\bwg$ decay process we are considering.
We should stress however, that, whereas the one-loop calculations
in the preceding section were explicit, here we are assuming that
we can incorporate our one-loop results into the general framework
of refs.~\cite{SCET2,SCET1}. This needs to be verified explicitly.

\subsection{The Matrix Element \boldmath{$F_\mu$}}

In this section we consider the resummation of the large
logarithms in the amplitude $F_\mu$. We start by separating the
total expression for $T^{(1)}$ into a contribution from the
electromagnetic and weak currents, including the corresponding
contributions due to wave function renormalization (since the
contribution from the box diagram is zero we do not discuss it
anymore in this section):
\begin{equation}
T^{(1)}=T^{(1)\,J_{\,\textrm{em}}}+T^{(1)\,J_{\,\textrm{W}}}\,
\end{equation}
where
\begin{eqnarray}\label{eq:t1em}
T^{(1)\,J_{\,\textrm{em}}}&=&T^{(1)\,\textrm{em}}+ \frac12
T^{(1)\,\textrm{wfc}}+T^{(1)\,\bar u\textrm{wf}}\nonumber\\ &=&
\frac{\as C_F}{4\pi}T^{(0)}(\tp)
  \left[\frac32\log \frac{2\tilde{k}\cdot q}{\mu^2_F}
  - \frac92 \right]\,,
\end{eqnarray}
and
\begin{eqnarray}\label{eq:t1wk}
T^{(1)\,J_{\,\textrm{W}}}&=&T^{(1)\,\textrm{wk}}+ \frac12
T^{(1)\,\textrm{wfc}}+T^{(1)\,b\textrm{wf}}\nonumber\\ &=&
\frac{\as C_F}{4\pi} T^{(0)}
   \Bigg[\frac12\log\frac{2\tilde k\cdot q}{\mu_F^2}+\frac12 \log\frac{m_b^2}{\mu_F^2}
   +\frac{1}{2}\log^2\frac{m_b^2}{\mu_F^2}
      \\
&& \hspace{-1in}
-\frac{\sqrt{2}q_-}{m_b-\sqrt{2}q_-}\log\frac{\sqrt{2}q_-}{m_b}
+\log^2\frac{m_b}{\sqrt{2}\tp}
      +2\log\frac{\mu_F^2}{\sqrt{2}q_-m_b} \log\frac{m_b}{\sqrt{2}\tp}
      +2\log\frac{m_b}{\sqrt{2}\tp}\nonumber\\
&&\hspace{-0.5in}  +\log^2\frac{m_b}{\sqrt{2}q_-}
      +2{\rm Li}_2\left(1-\frac{\sqrt{2}q_-}{m_b}\right)
      + 4 {\rm Li}_2\left(1-\frac{m_b}{\sqrt{2}q_-}\right)
      -\frac{\pi^2}{4}-\frac52\Bigg]\nonumber\,.
\end{eqnarray}
From eq.\,(\ref{eq:t1em}) we readily see that if we choose the
factorization scale $\mu_F^2$ to be of $O(q\cdot\tilde
k)=O(m_b\lqcd)$, then there are no large logarithms in
$T^{(1)\,J_{\,\textrm{em}}}$. On the other hand, this is not the
case for $T^{(1)\,J_{\,\textrm{W}}}$.

We focus therefore on the contribution from the weak vertex, which
involves a transition of the $b$-quark with momentum $p-k$ to a
$u$-quark with momentum $q-k$:
\begin{equation} \label{eq:wkextstate}
\langle u(q-k) | \bar{u}\gamma^\mu(1-\gamma_5)b | b(p-k)
\rangle\,.
\end{equation}
From eq.\,(\ref{eq:t1wk}) we see that even at one-loop order the
weak transition vertex has large double and single logarithms, and
it is these logarithms which we attempt to resum using the
techniques of the SCET~\cite{SCET2,SCET1}. Thus we are considering
the set of diagrams, where an arbitrary number of gluons is
exchanged between the incoming heavy quark and the collinear light
quark that links the weak and electromagnetic vertices. The
effective theory is designed to describe processes with heavy
quarks and (almost light-like) light quarks. Fluctuations below
some scale $\mu^2~(\ll m_b^2)$ are described in terms of the
effective theory and contributions from higher momenta are
obtained by perturbatively matching onto QCD and absorbed into
coefficient functions. In addition to heavy quarks, the effective
theory contains two kinds of light-quark and gluon fields. Using
light-cone coordinates (writing an arbitrary momentum $p$ as
$p=(p_+,p_-,\pv{p})$), the effective light-quark and gluon fields
are either soft $l\propto m_b(\lambda^2,\lambda^2,\lambda^2)$ or
collinear $l\propto m_b(\lambda^2,1,\lambda)$, with $\lambda$ a
small expansion parameter. In our case, the collinear quark
carries momentum $q-k$, and thus $\lambda=O(\sqrt{\lqcd/m_b})$. We
are interested in the leading-twist contribution to the matrix
element, and therefore in the SCET at leading order in $\lambda$
as discussed in ref.~\cite{SCET1} (for an extension to higher
orders in $\lambda$, see ref.~\cite{higher}).

The weak current in eq.~(\ref{eq:wkextstate})
can be matched onto the operators $O_i$ of the effective theory:
\begin{equation}
\bar{u}\gamma^\mu(1-\gamma_5)b = \sum_i C_i(\mu) O_i^\mu(\mu).
\end{equation}
The $C_i$ are Wilson coefficients which describe the physics above
$\mu$ and depend only on the hard scales such as $m_b$ and $q_-$.
The dependence of the Wilson coefficients, $C_i$, on $\mu$ cancels
the one which is implicit in the definition of the operators. The
large logarithms contained in the coefficient functions can be
resummed using renormalization group equations. As discussed
below, this means that the Wilson coefficients will receive
contributions from all orders of perturbation theory.

For the process which we are considering, the $\bwg$ decay, only
two SCET operators are relevant $O_3$ and $O_6$, which correspond
to the coefficient functions $C_3$ and $C_6$ in eq.\,(27) of
ref.~\cite{SCET1}. These two operators have the same Wilson
Coefficient $C(\mu_F)\equiv C_3(\mu_F)=C_6(\mu_F)$ (in fact, only
the difference $O_3-O_6$ is relevant for the left-handed current
in this process). Thus we need the renormalization group equation
for a single coefficient function.

In Ref.~\cite{SCET1} one-loop results were obtained for the weak
current. When convoluted with the lowest order wave function and
rewritten in our notation, these results are:
\begin{eqnarray} \label{eq:ffscet}
F^{(0)}_\mu+F^{J_{\textrm{W}}}_\mu&\equiv&F^{(0)}_\mu+
F^{(1)\,\textrm{wk}}_\mu +\frac12
F^{(1)\,\textrm{wfc}}_\mu+F^{(1)\,b\textrm{wf}}_\mu\nonumber\\
  &&\hspace{-0.7in}= C(\mu_F)\ F^{(0)}_\mu\left[1+ \frac{\as(\mu_F) C_F}{4\pi}
  [L_s^{\rm wk} + L_s^{b{\rm wf}} + L_c^{\rm wk}+ L_c^{\rm wfc}]\right]\,,
\end{eqnarray}
where
\begin{eqnarray}
L_s^{\rm wk}(\mu_F) &=&
  -(N_\varepsilon^{UV})^2 - 2N_\varepsilon^{UV} \log\frac{\mu_F}{\sqrt{2}k_+}
  \label{eq:lstart}\\
&&\quad  -2\log^2\frac{\mu_F}{\sqrt{2}k_+} - \frac{3\pi^2}{4}
\nonumber\\ L_s^{b{\rm wf}}(\mu_F) &=&
N_\varepsilon^{UV}+2\log\frac{\mu_F}{\sqrt{2}k_+}\label{eq:lsend}\\
L_c^{\rm wk}(\mu_F) &=&
  2(N_\varepsilon^{UV})^2 + 2N_\varepsilon^{UV} + 2N_\varepsilon^{UV}
  \log\frac{\mu_F^2}{2(q\cdot k)}\\
&&\quad  +\log^2\frac{\mu_F^2}{2(q\cdot k)}+2\log\frac{\mu_F^2}{2(q\cdot k)}
  +4-\frac{\pi^2}{6} \nonumber\\
L_c^{\rm wfc}(\mu_F) &=&
  -\frac{1}{2} N_\varepsilon^{UV}-\frac{1}{2}
  \log\frac{\mu_F^2}{2(q\cdot k)}-\frac{1}{2}\,. \label{eq:lend}
\end{eqnarray}
$L_s$ ($L_c$) comes from the exchange of a soft (collinear) gluon,
and the superscripts ``wk'', ``$b$wf'' and ``wfc'' denote
respectively the one-loop contributions from the weak vertex, the
wave function renormalization for the heavy quark and (half of)
that of the collinear light-quark. One might have also expected
two additional terms on the right-hand side of
eq.~(\ref{eq:ffscet}), $L_s^{\rm wfc}$ and $L_c^{b{\rm wf}}$.
However $L_s^{\rm wfc}$ and $L_c^{b{\rm wf}}$ vanish in the
effective theory because of the decoupling of soft gluons and
collinear quarks and of collinear gluons and heavy quarks
respectively.

At this point we should point out that the result for $L_s^{b{\rm
wf}}$ differs from that in eq.\,(35) of ref.~\cite{SCET1} by a
constant term. This is due to our choice for the definition of
$Z_2$ when the infra-red divergences are regulated by taking the
quark off-shell (we define $Z_2$ through the derivative of the
self-energy diagram with respect to the external momentum as in
eqs.\,(\ref{eq:z2defq}) and (\ref{eq:z2defQ})\,). The difference
cancels in the matching, as long as one consistently uses the same
definition in the effective theory and QCD, leading to the same
result for $T^{(1)}$.

The matching of SCET onto QCD at the scale $\mu=m_b$ yields the
Wilson coefficient at this scale:
\begin{eqnarray}\label{eq:matchingc3}
C(m_b)&=&1-\frac{\as(m_b)C_F}{4\pi}\\ &&
\times\left[2\log^2\frac{y}{x}+2{\rm
Li}_2\left(1-\frac{y}{x}\right)+\frac{3y-2x}{x-y}\log\frac{y}{x}
          +\frac{\pi^2}{12}+6\right]\,,\nonumber
\end{eqnarray}
where $x$ and $y$ have been defined in eq.~(\ref{eq:xyzdef}).

We now consider the renormalization group equation for $C(\mu)$
and study its scale dependence. From the one-loop results in
eqs.~(\ref{eq:lstart})-(\ref{eq:lend}), one obtains the evolution
equation for the Wilson coefficient function~\cite{SCET1}
\begin{equation}
\mu\frac{d}{d\mu}C(\mu)=\gamma(\mu) C(\mu)\,,
\end{equation}
where the first two terms of the anomalous dimension $\gamma$ are
given by
\begin{eqnarray}
\gamma_{LO}&=&-\frac{\as(\mu)C_F}{\pi}\log\frac{\mu}{\sqrt{2}q_-}\,\\
\gamma_{NLO}&=&-\frac{5\as(\mu)C_F}{4\pi}
  -2C_FB \frac{\as^2(\mu)}{4\pi^2}\log\frac{\mu}{\sqrt{2}q_-}\,
\end{eqnarray}
with $B= C_A(67/18-\pi^2/6)-5N_f/9$. The value of $B$ is deduced
in ref.~\cite{SCET1} by comparison with earlier work on inclusive
$B\to X_s\gamma$ and $B\to X_u \ell\bar{\nu}$ decays~\cite{KS,AR}.
The coefficient $B$ can (and should) be checked directly by a
two-loop computation within the SCET framework.

The homogeneous nature of the evolution equation leads to the
exponentiation of (Sudakov) logarithms:
\begin{equation}\label{eq:cmu}
C(\mu)=C(m_b)\exp\left[\frac{f_0(r)}{\as(m_b)}+f_1(r)\right]\,,
\end{equation}
where
\begin{eqnarray}
f_0(r)&=&-\frac{4\pi C_F}{\beta^2_0}\left[\frac{1}{r}-1+\log
r\right]\,,\\
f_1(r)&=&-\frac{C_F\beta_1}{\beta_0^3}\left[1-r+r\log
r-\frac{\log^2 r}{2}\right]\\ &&\quad
+\frac{C_F}{\beta_0}\left[\frac{5}{2}-2\log\frac{y}{x}\right]\log
r
       -\frac{2C_F B}{\beta_0^2}[r-1-\log r]\,,\nonumber
\end{eqnarray}
with
\begin{equation}
r=\as(\mu)/\as(m_b)\,,
\end{equation}
$\beta_0=11C_A/3 -2N_f/3$ and $\beta_1=34C_A^2/3-10C_A N_f/3 -2C_F
N_f$.

\begin{figure}[t]
\begin{center}
\includegraphics[width=9cm,angle=270]{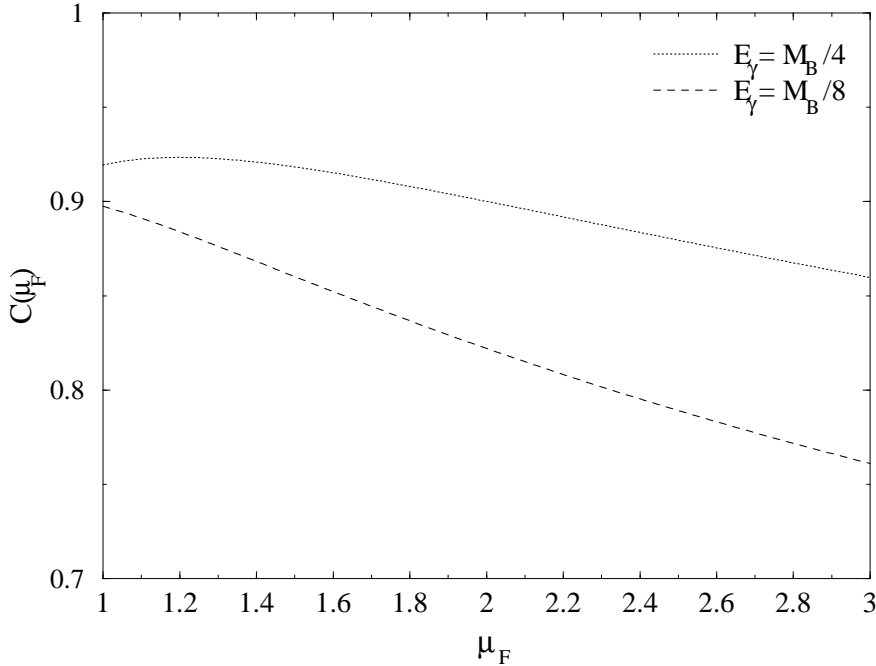}
\caption{Value of the Wilson coefficient $C$ at next-to-leading
order as a function of $\mu$, for two different values of the
photon energy: $\eg=M_B/4$ (dotted line) and $\eg=M_B/8$ (dashed
line).} \label{fig:cmu}
\end{center}
\end{figure}

In fig.~\ref{fig:cmu} the Wilson coefficient $C(\mu)$ is plotted
as a function of $\mu$ (at next-to-leading order as in
eq.~(\ref{eq:cmu})\,). At this order, $C$ is also a function of
the photon energy $\eg$, and for illustration we chose two
different values of $\eg$. We note that for $\mu$ of order
$\sqrt{m_b\lqcd}$, $C$ differs from 1 typically by 10-20\%.

The above equations allow us to resum the large logarithms present
in $T^{(1)}$. Before doing this however, we verify that our
results at one-loop order agree with those of ref.~\cite{SCET1}.
Expanding the exponential in eq.~(\ref{eq:cmu}) we obtain
\begin{eqnarray}\label{eq:runningc3}
C(\mu_F)&=&C(m_b)\times\\
 &&\hspace{-0.5in} \left[1+\frac{\as C_F}{4\pi} \left(
      -2\log^2\frac{\mu_F}{m_b}+4\log\frac{\sqrt{2}q_-}{m_b}\log\frac{\mu_F}{m_b}
      -5\log\frac{\mu_F}{m_b} \right) \right] + O(\as^2)\,.\nonumber
\end{eqnarray}
We now need to combine the one-loop term in
eq.~(\ref{eq:runningc3}) with the one-loop expression for the
coefficient function at the scale $m_b$ in
eq.~(\ref{eq:matchingc3}) and the remaining one-loop terms in
eq.~(\ref{eq:ffscet}), using the explicit expressions in
eqs.~(\ref{eq:lstart})-(\ref{eq:lend}). The dependence on the
factorization scale disappears as expected, and the result is
identical to our result for $F_\mu^{(0)}+F_\mu^{J_{\textrm{W}}}$,
see eqs~(\ref{eq:f0budef}), (\ref{eq:fse}), (\ref{eq:fbwf}) and
(\ref{eq:fwk}).

\subsection{The Distribution Amplitude and Hard-Scattering Kernel}

In the previous subsection, we have checked that the SCET
correctly reproduces the one-loop contribution of the weak current
to the matrix element. In this section we investigate whether the
contribution to the distribution amplitude can be understood in
the effective theory as well. We shall see that this is indeed the
case, allowing us to give an expression for the hard-scattering
kernel in which the large Sudakov logarithms are resummed.

Recall that the contribution to the distribution amplitude
corresponds to Feynman diagrams where the propagator of the
internal light quark is replaced by its eikonal approximation. In
the framework of SCET, this approximation is automatically
performed when the interaction of soft gluons and collinear quarks
is considered. We expect therefore that the contribution to the
distribution amplitude in our framework is equal to the exchange
of a soft gluon in the SCET framework.

When we compare our results for the contributions to
$\Phi^{(1)}\otimes T^{(0)}$ in eqs.~(\ref{eq:phiQ}) and
(\ref{eq:phiwk}) with those for $L_s^{b\textrm{wf}}$ and
$L_s^{\textrm{wk}}$ in eqs.~(\ref{eq:lstart}) and
(\ref{eq:lsend}), we find explicitly that:
\begin{equation}
\Phi^{(1) Q{\rm wf}}\otimes T^{(0)}=\frac{\as
C_F}{4\pi}T^{(0)}(k_+)\,L_s^{b{\rm wf}}\end{equation} and
\begin{equation}
\Phi^{(1) {\rm wk}} \otimes T^{(0)}=\frac{\as
C_F}{4\pi}T^{(0)}(k_+)\,L_s^{\rm wk}\,,
\end{equation}
for any value of the factorization scale. Both
$\Phi^{(1)\,\textrm{wfc}}$ and $L_s^{\textrm{wfc}}$ are zero. The
soft terms, the $L_s$'s, do therefore correspond to the terms
absorbed in the distribution amplitude and the hard-scattering
kernel contains the remaining terms, i.e. the collinear
contributions $L_c$ and the resummed logarithms in the Wilson
coefficient $C$.

\begin{figure}[t]
\begin{center}
\includegraphics[height=7cm]{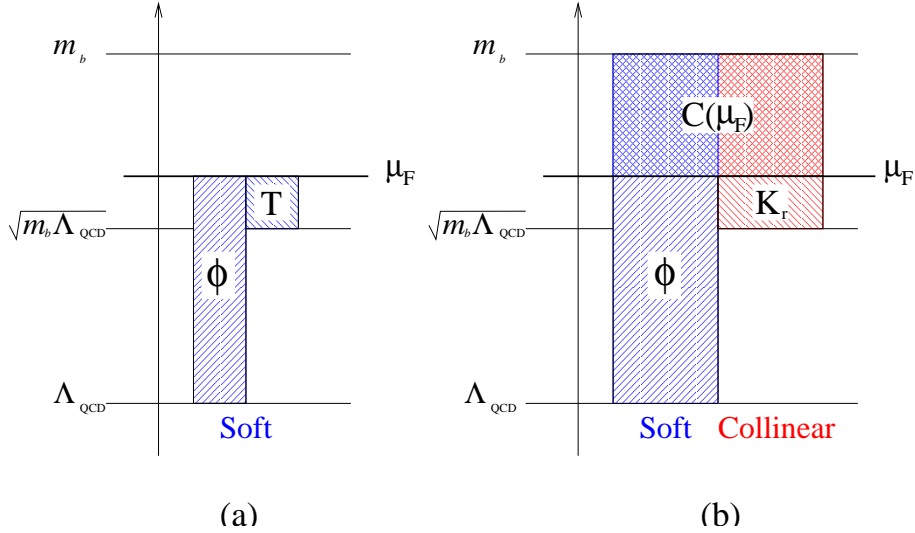}
\caption{Schematic representation of the large logarithms present
in the contributions to the matrix element in the case of (a) the
electromagnetic vertex and (b) the weak vertex. $M$ represents the
large scale, $M\simeq m_b$ and $M_B$.} \label{fig:scales}
\end{center}
\end{figure}

At this point it may be instructive to consider the various scales
which are present in the calculation, and we illustrate the
corresponding contributions to the matrix element in
Fig.~\ref{fig:scales}. Fig.~\ref{fig:scales}(a) illustrates the
situation for the electromagnetic vertex. The contributions from
scales between $\lqcd$ and $\sqrt{m_b\lqcd}$ are ``soft", in the
sense that all components of momenta are small and the diagrams
satisfy the SCET Feynman rules for soft gluons (although, as seen
in sec.~\ref{subsec:emvertex}, in the evaluation of diagrams the
components are not always uniformly small). The mass singularities
and remaining contributions from scales between $\lqcd$ and
$\sqrt{m_b\lqcd}$, are absorbed into the distribution amplitude,
so that hard-scattering kernel has no large logarithms for
$\mu_F=O(\sqrt{m_b\lqcd})$. The situation is more complicated for
the weak vertex in Fig.~\ref{fig:scales}(b). The matrix element
gets large logarithms from soft and collinear regions of phase
space. The soft contributions come from scales between $m_b$ and
$\lqcd$, and the collinear ones from $m_b$ down to
$\sqrt{m_b\lqcd}$. The distribution amplitude absorbs the soft
contributions below the factorization scale $\mu_F$ (including the
corresponding mass singularities). The hard-scattering kernel thus
contains large logarithms coming from the region between $m_b$ and
$\sqrt{m_b\lqcd}$ and it is these logarithms which we resum into
the Wilson Coefficient $C$. The hard-scattering kernel has a
factor of $C$, containing the resummed large logarithms and a
remaining one-loop term $K_r$ with no large logarithms.

The contribution from the weak current can then be factorized as
follows:
\begin{eqnarray}
F^{(0)}_\mu+F^{(1)\,J_{\rm W}}_\mu&=&C(\mu_F) F^{(0)}_\mu
 \left[1+\frac{\as(\mu_F) C_F}{4\pi} [L_s^{\rm wk} +
 L_s^{b{\rm wf}} + L_c^{\rm wk}+ L_c^{\rm wfc}](\mu_F)\right]
\nonumber\\
 &&\hspace{-1.35in}=\int \frac{d\tp}{2\pi} C(\mu_F)
 \left[1+\frac{\as(\mu_F) C_F}{4\pi} [L_c^{\rm wk}+
 L_c^{\rm wfc}](\tp;\mu_F)\right]T^{(0)}(\tp)
          \Phi(\tp;\mu_F)\,.
\end{eqnarray}
$\Phi$ includes the soft logarithms of $\mu_F^2/\tp^2$. The
hard-scattering kernel is the product of the coefficient $C$,
which collects the resummed (exponentiated) logarithms of $\mu_F$
over large scales ($m_b$, $q_-$), and of the ``collinear''
contributions from the $L_c$'s, i.e. the logarithms of
$\mu_F^2\sim 2(q\cdot \tilde{k})$. If the factorization scale is
set to $\mu_F^2=O(m_b\lqcd)$, the large logarithms of the matrix
element are either included in the (non-perturbative) distribution
amplitude or resummed in the coefficient $C$.

Finally we combine the results from the corrections to the weak
and electromagnetic currents. We expect that the Wilson
coefficient $C$, which resums the Sudakov logarithms, is
associated with the weak current in any diagram and thus
multiplies all the remaining contributions (the validity of this
expectation needs to be investigated at higher orders). We thus
obtain the one-loop factorized expression for the form factors,
including the resummation of leading and next-to-leading
logarithms:
\begin{equation}\label{eq:fvastart}
F_A(\eg)=F_V(\eg)= \int d\tp \Phi^B_+(\tp;\mu_F)\,
T(\tp,\eg;\mu_F)\,,
\end{equation}
where
\begin{eqnarray}
T(\tp,\eg;\mu_F)&=&C(\mu_F)\,\frac{f_B Q_u M_B}{2\sqrt{2}\eg}
\frac{1}{\tp}\times\nonumber\\ &&\hspace{0.5in} \left[1+\frac{\as
(\mu_F)\, C_F}{4\pi} K_t(\tp,\eg;\mu_F) \right]
\end{eqnarray}
and
\begin{equation}
K_t(\tp,\eg;\mu_F) =
\log^2\frac{2q_-\tp}{\mu_F^2}-\frac{\pi^2}{6}-1\,.
\label{eq:fvaend}\end{equation} $K_t$ is the sum of two
contributions: that from the electromagnetic current and $K_r$,
the contribution from the weak current which has not been included
in the resummation.

In summary, following the formalism in ref.~\cite{SCET1}, we have
resummed the leading and next-to-leading logarithms of the form
$\as^n\log^{n+1}(m_b/\sqrt{2}\tp)$ and
$\as^n\log^n(m_b/\sqrt{2}\tp)$. Next-to-next-to-leading
logarithms, of the form $\as^n\times$\\
$\log^{n-1}(m_b/\sqrt{2}\tp)$ have not been resummed here
(although we do present the one-loop term of the form
$\alpha_s\times$ constant in eqs.~(\ref{eq:fvastart}) --
(\ref{eq:fvaend})). These terms would require a three-loop
calculation in the effective theory~\cite{SCET1}.

\section{Phenomenological Study}\label{sec:phenom}

In this section we briefly investigate the phenomenological
consequence of the analysis presented above for the decay $B^+\to
e^+\nu\gamma$ (the discussion of the decay $B^+\to\mu^+\nu\gamma$
would be the same). The current experimental bounds on the
branching ratios are $\Gamma(B^+\to e^+\nu\gamma)/\Gamma_B <
2.0\times 10^{-4}$ and $\Gamma(B^+\to \mu^+\nu\gamma)/\Gamma_B <
5.2\times 10^{-5}$ (both at 90\,\% confidence level) \cite{PDG}.
$\Gamma_B$ is the total decay width of the $B$-meson. Previous
theoretical estimates of the branching ratios are typically in the
range $(1-5)\times 10^{-6}$~\cite{KPY,previous}.

The differential decay rate for the process $\bwg$ in terms of the
form-factors $F_V$ and $F_A$ is given by~\cite{KPY}:
\begin{eqnarray}
\frac{d^2\Gamma}{dE_e dE_\gamma} &=&\frac{\alpha G_F^2 |V_{ub}|^2
M_B^3}{16\pi^2}\, \Big\{
R[R^2+2S(S-R-1)+1][F^2_V+F_A^2](\eg)\nonumber\\ &&\qquad \qquad
-2R(1-R)(1+R-2S)F_V(\eg)F_A(\eg)
  \Big\}\,,
\end{eqnarray}
where $R=1-2\eg/M_B$, $S=2E_e/M_B$ and $E_\gamma$ and $E_e$ are
the energies of the photon and electron respectively. $R$ and $S$
satisfy $0\leq R\leq 1$ and $R\leq S\leq 1$. Integrating over the
electron's energy, we obtain the differential decay rate
\begin{equation}
\frac{d\Gamma}{d\eg}
  =\frac{\alpha G_F^2 |V_{ub}|^2 M_B^4}{48\pi^2} R(1-R)^3
  [F^2_V+F_A^2](\eg)\,.
\end{equation}

Our analysis of the form factors $F_V$ and $F_A$, is valid only if
the energy of the photon is large. We therefore introduce a lower
cutoff $E_\gamma^c$ and consider the integrated decay rate
\begin{equation}
\Gamma(\eg^c)=
  \int_{\eg^c}^{M_B/2}d\eg\, \frac{d\Gamma}{d\eg}\,.
\end{equation}
$\eg^c$ is formally of $O(M_B/2)$.

At leading order of perturbation theory, the two form factors are
given by
\begin{equation}
F_A^{LO} = F_V^{LO} =
  \frac{f_B Q_u M_B}{2\eg\lambda_B}=\frac{1}{1-R}\ \frac{f_B
  Q_u}{\lambda_B}\,,
\end{equation}
where $\lambda_B$ is the first inverse moment of the $B$-meson's
distribution amplitude,
\begin{equation}
\frac{\sqrt{2}}{\lambda_B}=\int_0^\infty \frac{d\tp}{\tp}
\Phi_+^B(\tp)\,. \label{eq:lambdabdef}\end{equation} Little is
known about this quantity. Below, we will use the estimate of
ref.~\cite{BBNS2}, $\lambda_B=350\pm 150$~MeV~\footnote{The
definition of $\lambda_B$ here is the same as in
ref.~\cite{BBNS2}, in spite of the different choice for the
normalization of $\phi_+^B$.}.

The integrated rate at leading order is
\begin{equation}
\Gamma^{LO}(\eg^{\rm c})=
  \frac{\alpha Q_u^2 G_F^2 |V_{ub}|^2}{96\pi^2}\,
  \frac{M_B^5\,f_B^2}{\lambda_B^2}\,R_c^2\left(1-\frac{2}{3}R_c\right)
\end{equation}
where $R_c=1-2\eg^c/M_B$. Dividing the integrated rate by the
experimentally measured total width ($\Gamma_B$), we obtain the
numerical LO estimate of the branching ratio ($B^{LO}(\eg^{\rm
c})\equiv\Gamma^{LO}(\eg^{\rm c})/\Gamma_B$)
\begin{eqnarray}
B^{LO}(\eg^{\rm c}) &=& 18.4\times10^{-6}
\left(\frac{|V_{ub}|}{3.6\times 10^{-3}}\right)^{\hspace{-2.5pt}2}
\left(\frac{f_B}{190\,\textrm{MeV}}\right)^{\hspace{-2.5pt}2}
\left(\frac{350\,\textrm{MeV}}{\lambda_B}\right)^{\hspace{-2.5pt}2}\nonumber\\
&&\hspace{1in}\times R_c^2\left(1-\frac{2R_c}{3}\right).
\label{eq:blo}\end{eqnarray} The central values for $|V_{ub}|$ and
$f_b$ were taken from refs.\,\cite{PDG} and \cite{sinead}
respectively. We therefore expect a fully integrated branching
ratio $\Gamma$ of a few $10^{-6}$, unless there is a significant
enhancement of the soft-photon region of phase space (which is
beyond the reach of our framework).

It may be useful to normalize $\Gamma(\eg^{\rm c})$ by the purely
leptonic decay rate $\Gamma(B^+\to \mu^+\nu)$,
\begin{equation}
\Gamma(B^+\to \mu^+\nu)=\frac{G_F^2|V_{ub}|^2 f_B^2 M_B
M_\mu^2}{8\pi} \left(1-\frac{M_\mu^2}{M_B^2}\right)\,,
\end{equation}
since it absorbs the dependence on two relatively poorly known
quantities: the $B$-meson decay constant, $f_B$, and the modulus
of the CKM matrix element $|V_{ub}|$. (For the central values of
$|V_{ub}|$ and $f_B$ in eq.~(\ref{eq:blo}) the branching fraction
for the mode $\mu\nu$ is about $3.7\times 10^{-7}$. The current
experimental upper bound is $ 2.1\times 10^{-5}$ at 90\%
confidence level~\cite{PDG}.)

The integrated rate for the decay $B^+\to e^+\nu\gamma$ is thus
(at leading order and neglecting terms of $O(M_\mu^2/M_B^2)$)
\begin{eqnarray}
\frac{\Gamma(\eg^{\rm c})}{\Gamma(B^+\to\mu^+\nu)}
 &=&\frac{\alpha Q_u^2}{12\pi}\frac{M_B^2}{M_\mu^2}
       \frac{M_B^2}{\lambda_B^2} R_c^2\left(1-\frac{2}{3}R_c\right)\\
&=&0.22\left(\frac{M_B}{\lambda_B}\right)^2
           R_c^2\left(1-\frac{2}{3}R_c\right)\nonumber\\
&=&50\,\left(\frac{.35\,\textrm{GeV}}{\lambda_B}\right)^2
R_c^2\left(1-\frac{2}{3}R_c\right)\,.\label{eq:gammas}
\end{eqnarray}
For the range of values of $\lambda_B$ given after
eq.~(\ref{eq:lambdabdef}), the product of the first two factors in
eq.~(\ref{eq:gammas}) varies in the range (25,\,150).

We now study how this result is modified at next-to-leading order
in $\as$. It is convenient to rewrite the result for the form
factors in eqs.\,(\ref{eq:fvastart}) -- (\ref{eq:fvaend}) in the
form,
\begin{eqnarray} \label{eq:nloff}
F_{A,V}&=&\frac{f_B Q_u M_B}{2\eg} C(\mu_F)
  \bigg[\left(1+\frac{\as(\mu_F) C_F}{4\pi}(a+L^2)\right)\frac{1}{\lambda_B}
  +\\
  &&\hspace{1.5in}2\,\frac{\as(\mu_F) C_F}{4\pi}\frac{L}{\lambda_B^{(1)}}
  +\frac{\as(\mu_F) C_F}{4\pi}\frac{1}{\lambda_B^{(2)}}\bigg]\,,
\nonumber\end{eqnarray} where $a=-\pi^2/6-1$,
\begin{equation}
L=\log\frac{\sqrt{2}q_-}{M_B}=\log(1-R)
\end{equation}
and
\begin{equation}\label{eq:lambdaBn}
\frac{\sqrt{2}}{\lambda_B^{(n)}}=\int_0^\infty \frac{d\tp}{\tp}
 \Phi^B_+(\tp) \log^n\frac{\sqrt{2}\tp M_B}{\mu_F^2}\,.
\end{equation}
The integrated decay rate, up to next-to-leading order, is:
\begin{eqnarray} \label{eq:integnlo}
\frac{\Gamma(\eg^c)}{\Gamma(B^+\to\mu^+\nu)}
 &=&\frac{\alpha Q_u^2}{6\pi}
  \frac{M_B^2}{M_\mu^2}
  \frac{M_B^2}{\lambda_B^2} \int_0^{R_c} dR\ R(1-R)\, C^2(\mu_F)\\
&&\hspace{-1in}  \times
 \left[1+\frac{\as(\mu_F) C_F}{4\pi}(a+L^2)
      +2\frac{\as(\mu_F) C_F}{4\pi}N_1 L
        +\frac{\as(\mu_F) C_F}{4\pi}N_2\right]^2\ ,
\nonumber
\end{eqnarray}
where the $N_k$'s are the ratios of inverse moments,
$N_k\equiv\lambda_B/\lambda_B^{(k)}$. $C(\mu_F)$ and $L$ depend
implicitly on the integration variable $R$.

We would like to estimate the size of the one-loop correction and
the effect of re-summing the leading and next-to-leading
logarithms. To this end it is necessary to re-expand the
coefficient $C(\mu_F)$ up to one-loop order:
\begin{eqnarray}
C(\mu_F)
 &=&1-\frac{\as C_F}{4\pi}
   \Bigg[2\log^2\frac{\mu_F}{M_B}-4\log(1-R)\log\frac{\mu_F}{M_B}
       +5\log\frac{\mu_F}{M_B}\nonumber\\
       &&\hspace{-0.4in}
       +2\log^2(1-R)+2{\rm Li}_2(R)+\frac{1-3R}{R}\log(1-R)+\frac{\pi^2}{12}+6
           \Bigg]\,.\label{eq:coeff1loop}
\end{eqnarray}
For any choice of input values for the cut-off $R_c$, the
factorization scale $\mu_F$, the large scale (which we have taken
to be  $M_B$ here) and the non-perturbative inverse moments
$\lambda_B^{(k)}$ we can perform the integration in
eq.~(\ref{eq:integnlo}) numerically. For illustration, in the
following we choose $\mu_F=1.25$~GeV, and $\lambda_B=0.35$~GeV (a
change of the choice of the value for $\lambda_B$, for fixed $N_1$
and $N_2$, amounts to a rescaling of the decay rate in the
discussion below).

\begin{figure}[ht]
\begin{center}
\includegraphics[width=9cm,angle=270]{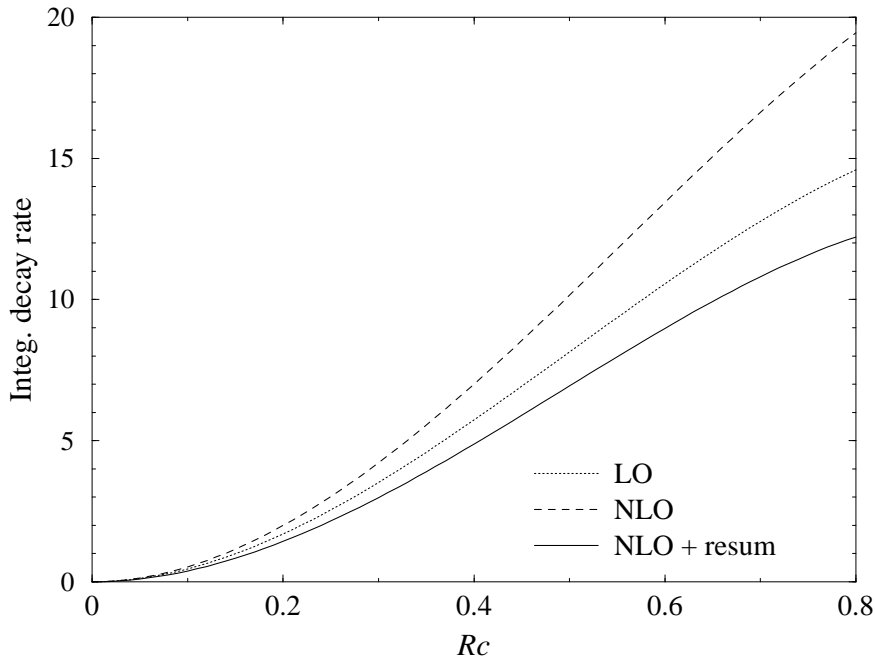}
\caption{Integrated decay rate, normalized by the purely leptonic
decay rate $B\to\mu\nu$, as a function of the cut-off
$R_c=1-2\eg^c$, when we neglect the NLO contributions with small
logarithms ($\alpha_s \times L^k$). The factorization scale is set
to $\mu_F=1.25$ GeV and the large scale is taken to be
$M_B=5.28$\,GeV. `LO'' and `NLO'' denote the leading and
next-to-leading results for the form factors, while `NLO+resum''
denotes the result obtained after resumming the leading and
next-to-leading logarithms. \label{fig:spectrum}}
\end{center}
\end{figure}

We now investigate how well we can evaluate the higher order
corrections and how large they are. As a first estimate of the
difference between the integrated decay rates calculated at LO and
at NLO, we neglect the ``small logarithms'', $\alpha_s \times L^k$
($k=0,1,2$). We will include these terms below, but it should be
noted that in the summation formalism which we have used in
sec.~\ref{sec:resum}, they are of the same order as terms which
have been neglected. From eq.~(\ref{eq:integnlo}), we see that now
the NLO decay rate becomes independent of the inverse moments
$\lambda_B^{(k)}$. The result is plotted in
figure~\ref{fig:spectrum}. In this figure we have also included
the one-loop result without resummation, replacing the Wilson
coefficient $C$ by its one-loop expression (\ref{eq:coeff1loop})
in eq.~(\ref{eq:integnlo}) and neglecting all the terms of order
$\alpha_s \times L^k$ [and taking $\alpha_s=\alpha_s(M_B)$ in the
resulting one-loop expression]. The figure shows the dependence of
the integrated decay rate on the cut-off and demonstrates that the
difference between the LO and (the resummed) NLO results is
typically of the order of about 25\%.

For the remainder of this section we include the one-loop terms in
the square parentheses in eq.~(\ref{eq:integnlo}), and investigate
the dependence of the decay rate on the small logarithm $L$ and on
the inverse moments $\lambda_B^{(k)}$. Formally all the
$\lambda_B^{(k)}$ are of the same order as $\lambda_B$, and thus
the $N_k$'s are of $O(1)$. They can however, take significantly
different values depending on the model used for the distribution
amplitude. This is illustrated by the following two models for the
$B$-meson distribution amplitude at the scale $\mu_F$:
\begin{eqnarray}
\Phi_+^{B\,I}(\tp)&=&\frac{2\tp}{\lambda_0^2}\exp\left(-\frac{\sqrt{2}\,\tp}{\lambda_0}\right)\,,
\\ \frac{1}{\lambda_B}& =& \frac{1}{\lambda_0}\\
\frac{1}{\lambda_B^{(1)}}
  &=&\frac{1}{\lambda_0}\left[\log\frac{\lambda_0 M_B}{\mu_F^2}-\gamma\right]\\
\frac{1}{\lambda_B^{(2)}}
  &=&\frac{1}{\lambda_0}\left[
 \left(\log\frac{\lambda_0 M_B}{\mu_F^2}-\gamma\right)^2+\frac{\pi^2}{6}\right]
\end{eqnarray}and
\begin{eqnarray}
\Phi_+^{B\,II}(\tp)&=&\frac{\sqrt{2}K\tp}{(\tp+K\lambda_0)^2}\,,\\
\frac{1}{\lambda_B}&=&\frac{1}{\lambda_0}\\
\frac{1}{\lambda_B^{(1)}}
  &=&\frac{1}{\lambda_0}\log\frac{\sqrt{2}\lambda_0M_B K}{\mu_F^2}\\
\frac{1}{\lambda_B^{(2)}}
  &=&\frac{1}{\lambda_0}\left[
 \log^2\frac{\sqrt{2}\lambda_0 M_B
 K}{\mu_F^2}+\frac{\pi^2}{3}\right]\,.
\end{eqnarray}
For instance, taking $\lambda_B=\lambda_0=0.35$ GeV and $K=0.3$ we
obtain the following values of the ratios of inverse moments
$N_k=\lambda_B/\lambda_B^{(k)}$:
\begin{equation}
I: N_1=-0.04\,,\ N_2=1.81\,,\qquad II:N_1=-0.69\,,\ N_2=3.77\,.
\end{equation}

The first model was proposed in ref.~\cite{GN}, based on QCD sum
rules. We propose the second one as an example of a distribution
amplitude with well defined inverse moments, but divergent
positive moments -- as expected from the one-loop study of the
renormalization properties of $\Phi^B$ above and in
ref.~\cite{GN}.

\begin{figure}[ht]
\begin{center}
\mbox{\includegraphics[width=6cm]{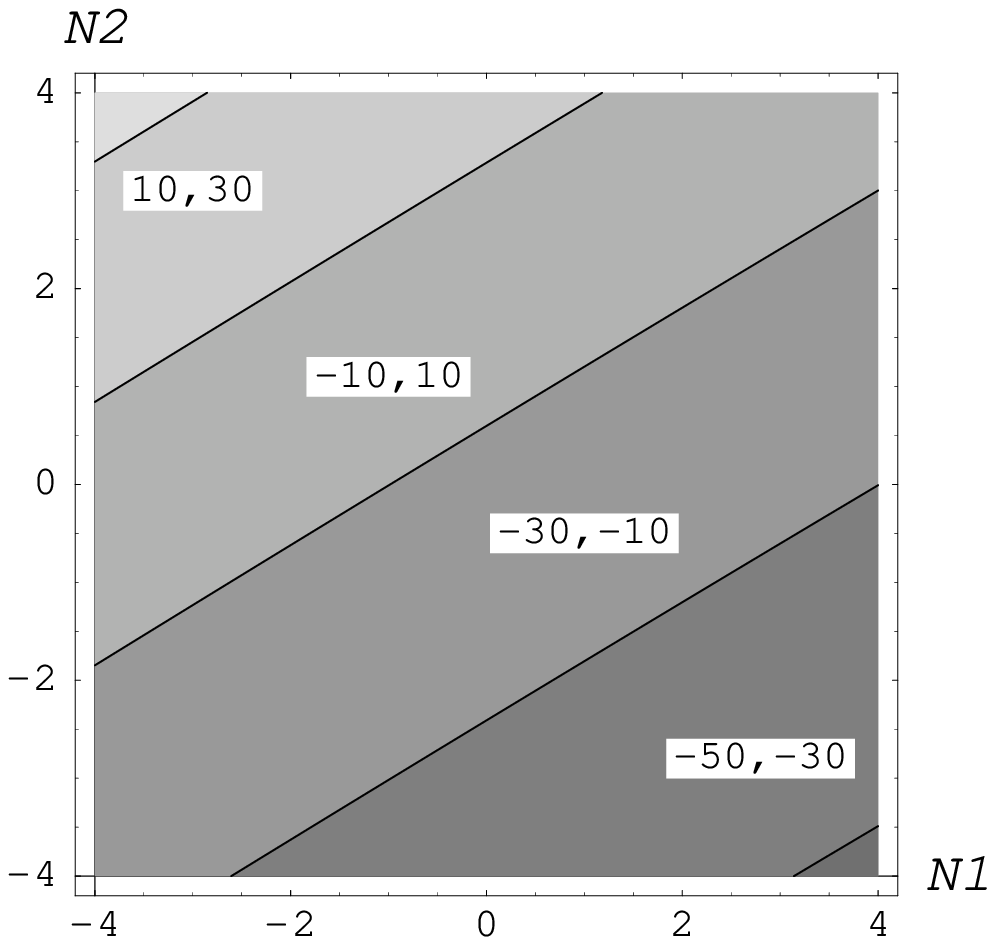}
\includegraphics[width=6cm]{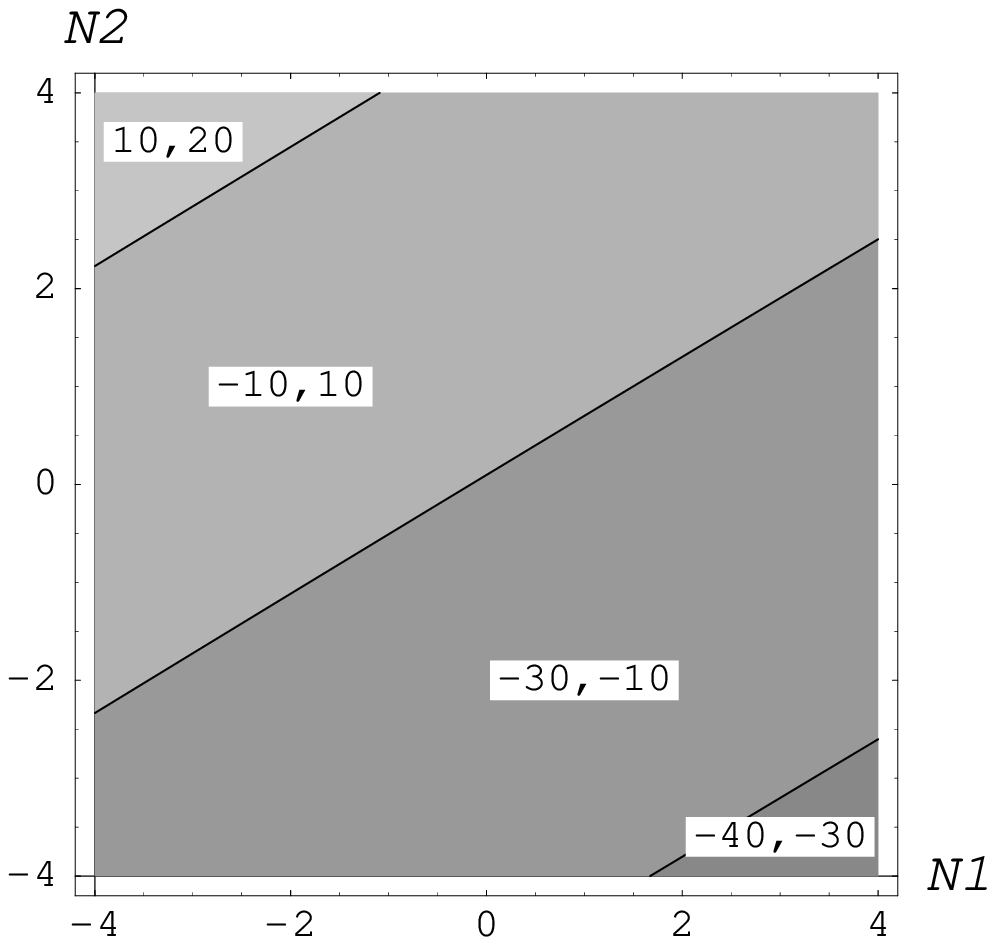}}\\ %
\mbox{\includegraphics[width=6cm]{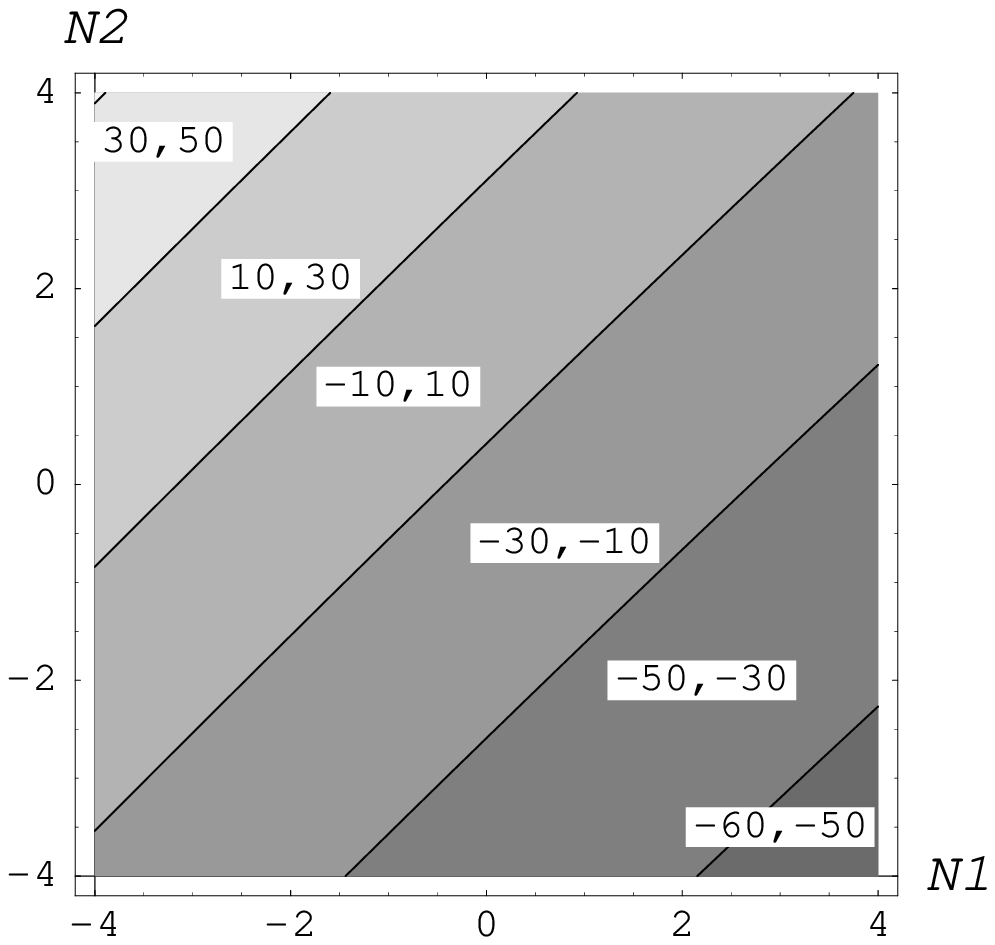}
\includegraphics[width=6cm]{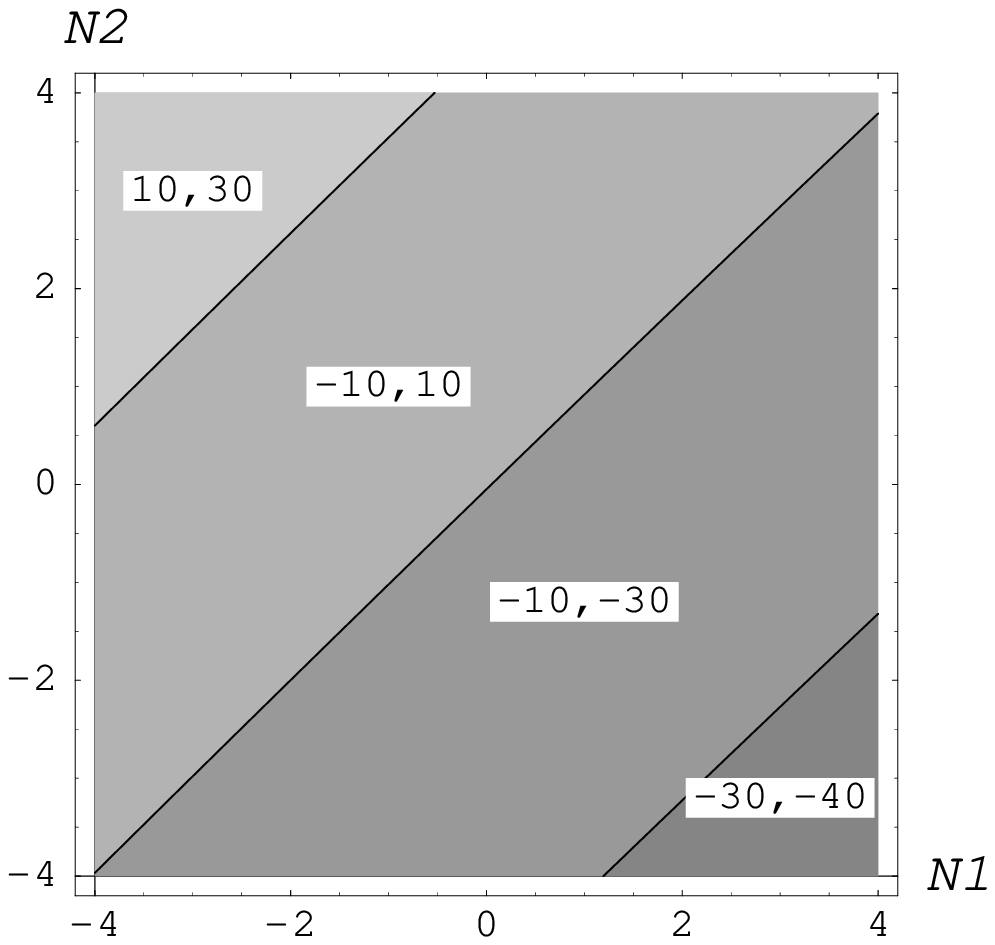}}
\caption{Difference between LO and NLO integrated decay rates as a
function of the two ratios of inverse moments $N_1$ and $N_2$, for
$R_c=0.4$ (upper line) and $R_c=0.6$ (lower line). In each case,
the plot on the left hand-side corresponds to the NLO resummed
expression (\ref{eq:integnlo}), and the plot on the right
hand-side to the one-loop expression [with $C$ expanded according
to eq.~(\ref{eq:coeff1loop})]. \label{fig:percent}}
\end{center}
\end{figure}

In figure~\ref{fig:percent} we plot  the difference (in percent)
between the LO results and the NLO ones, as a function of the two
ratios of inverse moments $N_1$ and $N_2$ for $-4\le N_{1,2}\le
4$. Each band in the figure corresponds to a range of 20\% and we
present the values at the corners ($N_{1,2}=\pm\,4$) in
table~\ref{tab:percent}. The one-loop corrections are typically
some (low) number of tens of percent and are seen to be sensitive
to the values of the ratios of inverse moments, $N_{1,2}$. Such a
significant variation of the rates with $N_{1,2}$ was to be
expected from an inspection of the integrand in
eq.~(\ref{eq:integnlo}). The relative size of the one-loop
corrections is also sensitive to the choice of the cut-off $R_c$.

\begin{table}[h]
\begin{center}
\begin{tabular}{|c|cc|cc|}
\hline & \multicolumn{2}{|c|}{$R_c=0.4$} &
\multicolumn{2}{|c|}{$R_c=0.6$}\\ $(N_1,N_2)$ & Resummed & 1-loop
only & Resummed & 1-loop only\\ \hline (-4,-4)& -24.6 \% & -16.8
\% & -13.2 \% & -10.1 \%\\ (4,-4) & -52.7 \% & -35.0 \% & -58.6 \%
& -39.5 \%\\ (-4,4) &  36.0 \% & 18.3 \% &  51.0 \% & 26.2 \%\\
(4,4)  & -2.8 \% & -3.7 \% & -11.7 \% & -9.1\%\\ \hline
\end{tabular}
\caption{Percentage difference between LO and NLO integrated decay
rates for various values of the two ratios of inverse moments
$N_1$ and $N_2$, for $R_c=0.4$ and $R_c=0.6$. In each case, we
give the percentage of difference from the LO results in the case
of the NLO resummed expression (\ref{eq:integnlo}), and the
one-loop expression [with $C$ expanded according to
eq.~(\ref{eq:coeff1loop})]. \label{tab:percent}}
\end{center}
\end{table}

Even from our simple exploratory study it is clear that $|V_{ub}|
f_B/\lambda_B$ could only be extracted from the integrated rate of
the radiative decay $B\to\gamma\ell\nu$ if we had precise
information of the $B$-meson's distribution amplitude (and in
particular its shape) from an independent process. Such knowledge
is necessary to estimate the potentially significant contribution
of the inverse moments $\lambda_B^{(1)}$ and $\lambda_B^{(2)}$ to
the NLO prediction.

We should stress again that the resummed formulae in
eqs.\,(\ref{eq:nloff}) and (\ref{eq:integnlo}) are not complete in
that we have not resummed the NNLO logarithms, which at one-loop
order are the terms which are proportional to $\alpha_s$ without
any large logarithms.

An interesting question, but one which is beyond the scope of the
present study, is whether it would be possible to invert the above
procedure and determine the important features of the $B$-meson's
distribution amplitude from the measured (in future) decay
distribution with sufficient precision for use in calculations of
decay rates of other processes. The $\bwg$ radiative decay, with
no hadrons other than the $B$-meson, seems to be a particularly
appropriate choice for such a determination of the distribution
amplitude.

\section{Conclusions}\label{sec:concs}

In this paper, we have studied the radiative decay $\bwg$ at
one-loop order in the framework of QCD factorization. We have
explicitly verified that factorization is valid at this order and
in eqs.\,(\ref{eq:aux}) and (\ref{eq:t1oneloop}) we present the
result for the hard-scattering amplitude. The non-perturbative
physics is contained in the $B$-meson's distribution amplitude,
which is defined on the light-cone (i.e. $z_+$ and $z_\perp$ are
set to zero in eq.\,(\ref{eq:Bmatrixelem})\,) and, at least at
this order of perturbation theory, there is no need to
re-formulate the factorization formalism in terms of wave
functions which depend also on the transverse momenta. On this
point we disagree with earlier claims\,\cite{KPY}.

In ref.\,\cite{KPY} the perturbative expansion of the light-cone
Bethe Salpeter wave function is not included in the matching
procedure and in addition, the wave function is redefined by a
perturbative factor $f(k_+, k_\perp)$ in order to modify the
evolution behaviour of the wave function (see eq.(26) in
ref.\,\cite{KPY}); the hard scattering kernel therefore is
redefined by a factor of $1/f$. This corresponds to trying to
absorb (at least part of) the contributions from
$\Phi^{(1)\,\textrm{box}}\otimes T^{(0)}$ (which depends on the
external transverse momentum) and $\Phi^{(1)\,\textrm{wk}}\otimes
T^{(0)}$ into $T^{(1)}$ by redefining the wave function. However,
we stress that such a redefinition is performed perturbatively and
that the non-perturbative physics is contained in the distribution
amplitude (\ref{eq:Bmatrixelem}).

The renormalization-group properties of the distribution amplitude
of the $B$-meson are very different from those of light mesons and
cannot be obtained from the positive moments~\cite{GN}. This makes
the determination of the distribution amplitude using standard
non-perturbative techniques, such as lattice simulations or QCD
sum rules, considerably more difficult. On the other-hand, one
should explore further the extent to which one might use future
experimental data on the photon energy distribution in $\bwg$
radiative decays to determine the properties of the distribution
amplitude which are needed in the phenomenology of $B$ decays to
two light-mesons.

The resulting one-loop hard-scattering kernel contains large
double and single logarithms of the ratio $m_b/\tp$, due to
Sudakov effects at the weak $b\to u$ vertex. These logarithms are
precisely those which one obtains from the Soft-Collinear
Effective Theory (SCET)~\cite{SCET1} (see
refs.~\cite{KPY,KS,wilsonline} and references therein for
discussions of the resummation of Sudakov logarithms using the
Wilson Line formalism). Assuming that this is also the case at
higher orders, we can resum the large logarithms using the
formalism developed in ref.~\cite{SCET1}. Moreover, we have been
able to match the SCET diagrams involving soft gluons with the
contributions to the distribution amplitude. Our phenomenological
study shows that the corrections are expected to be significant,
up to 50\% or so depending on the values of the parameters
$\lambda_B^{(n)}$ defined in eq.~(\ref{eq:lambdaBn}). Indeed, as
mentioned above, one can envisage in principle using $\bwg$ decays
to determine these parameters and to use them in calculating
predictions for other processes, such as two-body non-leptonic
$B$-decays.

An important motivation for this study is the need to develop the
QCD factorization formalism for two-body non-leptonic $B$-decays.
In particular, it is necessary to understand the \textit{hard
spectator interactions} (the term on the second line of
eq.\,(\ref{fff})\,) beyond the tree-level, and the corresponding
loop corrections share the key features of the calculations
described in this paper. These include the dependence on the three
scales and the need to control the Sudakov large logarithms.

Our conclusions for the development and exploitation of the QCD
Factorization formalism are very optimistic but a considerable
amount of work still needs to be done. The explicit demonstration
of factorization at one-loop order in this paper, needs to be
extended to higher orders. Most probably techniques such as those
incorporated into the SCET will be very useful in this context.
These studies have also to be applied to the decays of $B$-mesons
into two light mesons, so that the large (and growing) amount of
experimental data, predominantly from the $B$-factories, can be
analyzed in terms of the fundamental parameters of QCD (in
particular the CKM-matrix elements) and provide an understanding
of CP-violation in the quark sector.

\section*{Acknowledgements}

We warmly thank Grisha Korchemsky for many valuable and
instructive discussions. We are also grateful to M.~Beneke,
E.~Gardi and D.~Pirjol for helpful comments.

SDG thanks Prof.~G.~Altarelli and the Theory Division at CERN for
their hospitality during the completion of this work. This
research has been partially supported by PPARC, through grants
PPA/G/O/1998/00525 and PPA/G/S/1998/00530.

\end{document}